\documentclass[manuscript,screen]{acmart}

\usepackage[T1]{fontenc}
\usepackage{xspace}
\usepackage{color}
\usepackage{xcolor}
\usepackage{soul} 
\usepackage{enumitem} 
\usepackage{microtype} 
\PassOptionsToPackage{warn}{textcomp} 
\usepackage{textcomp} 
\usepackage{mathptmx} 
\usepackage{tabu} 
\usepackage{booktabs} 
\usepackage{subfiles}

\usepackage{amsmath}
\usepackage{algorithm,algorithmic}

\usepackage{MnSymbol}

\definecolor{brickred}{rgb}{0.8, 0.1, 0.1}

\sethlcolor{Goldenrod}
\newcommand{\xiaoyu}[1]{{\color{purple}[Xiaoyu\@: #1]}}
 
\newcommand{\name}{ConceptScope}

\newcommand{\algcom}[1]{{\color{gray} #1}}

\AtBeginDocument{%
  \providecommand\BibTeX{{%
    \normalfont B\kern-0.5em{\scshape i\kern-0.25em b}\kern-0.8em\TeX}}}

\setcopyright{acmcopyright}

\copyrightyear{2021}
\acmYear{2021}
\setcopyright{acmcopyright}\acmConference[CHI '21]{CHI Conference on Human
Factors in Computing Systems}{May 8--13, 2021}{Yokohama, Japan}
\acmBooktitle{CHI Conference on Human Factors in Computing Systems (CHI '21),
May 8--13, 2021, Yokohama, Japan}
\acmPrice{15.00}
\acmDOI{10.1145/3411764.3445396}
\acmISBN{978-1-4503-8096-6/21/05}

\begin{document}

\title[\name]{\name: Organizing and Visualizing Knowledge in Documents
based on Domain Ontology}

\author{Xiaoyu Zhang}
\email{xybzhang@ucdavis.edu}
\affiliation{%
  \institution{Department of Computer Science,
               University of California, Davis}
   \streetaddress{1 Shields Ave}
   \city{Davis}
   \state{California}
   \postcode{95616}
}

\author{Senthil Chandrasegaran}
\email{r.s.k.chandrasegaran@tudelft.nl}
\affiliation{%
  \institution{Faculty of Industrial Design Engineering,
               Delft University of Technology}
  \streetaddress{Landbergstraat 15}
  \city{2628 CE Delft, The Netherlands}
}

\author{Kwan-Liu Ma}
\email{klma@ucdavis.edu}
\affiliation{%
  \institution{Department of Computer Science,
               University of California, Davis}
  \streetaddress{1 Shields Ave}
  \city{Davis}
  \state{California}
  \postcode{95616}
}


\begin{abstract}
  Current text visualization techniques typically provide overviews of document content and structure using intrinsic properties such as term frequencies, co-occurrences, and sentence structures.
Such visualizations lack conceptual overviews incorporating domain-relevant knowledge, needed when examining documents such as research articles or technical reports.
To address this shortcoming, we present ConceptScope, a technique that utilizes a domain ontology to represent the conceptual relationships in a document in the form of a Bubble Treemap visualization.
Multiple coordinated views of document structure and concept hierarchy with text overviews further aid document analysis.
ConceptScope facilitates exploration and comparison of single and multiple documents respectively.
We demonstrate ConceptScope by visualizing research articles and transcripts of technical presentations in computer science.
In a comparative study with DocuBurst, a popular document visualization tool, ConceptScope was found to be more informative in exploring and comparing domain-specific documents, but less so when it came to documents that spanned multiple disciplines.

\end{abstract}



\maketitle


\section{Introduction}
\vspace{-2mm}
Text visualization techniques have evolved as a response to the virtual
explosion of text data available online in the last few decades.
Specifically, they aim to provide a visual overview---what digital
humanities now call ``distant reading''~\cite{Moretti2005graphs}---of
large documents or large collections of documents, and help the
researcher, investigator, or analyst find text patterns within and
between documents (e.g.~\cite{Stasko2008jigsaw}).
Most of these visualization techniques are domain-independent and do not
provide a knowledge-based overview of documents.
There have been approaches to provide a visual overview of the semantic
content of documents (e.g.~\cite{Collins2009docuburst}).
Such approaches have typically looked to lexical hypernymy (is-a
relationships) to provide a conceptual overview of the text.

However, when examining domain-specific documents such as research
papers, medical reports, or legal documents, it is necessary to examine
the documents from the point of view of that specific domain.
For instance, when examining a research paper in computer science, a
computer science researcher may be interested in whether the paper
concerns a general overview of a subject, such as ``computer graphics'',
or concerns more specific concepts such as ``infographics'' or ``TreeMap
visualizations''. 
Similarly, the researcher may want to compare papers that appear in the
same conference session to see the similarities and differences that may
exist between the papers.
In such scenarios, the overview visualizations should also represent
the computer science domain and how the knowledge is structured in the
domain.

While approaches such as topic modeling can provide a bottom-up
categorization or thematic separation of a document's text, domain
knowledge is often organized formally by experts in the corresponding
domains using Ontologies.
An ontology, defined as an ``explicit specification of a
conceptualization''~\cite[p.\ 199]{Gruber1993translation}, is a
widely-accepted way in which domain knowledge is formally represented.
A knowledge-based overview of a document that uses as a reference the
corresponding domain ontology can thus provide a conceptual overview for
the domain expert.
Such a view can also be used structurally to help the expert compare two
or more documents based on the concepts they cover.

In order to aid document examination from the viewpoint of a specific
domain, we present \name, a text visualization technique that provides a
domain-specific overview by referring to a relevant ontology to infer
the conceptual structure of the document(s) being examined.
\name\ uses a Bubble Treemap view~\cite{Gortler2017bubble} to represent
concept hierarchies, highlighting concepts from the ontology that exist
within the document and their relationships with other concepts in the
document, as well as key ``parent'' concepts in the Ontology.
Each concept ``bubble'' is also populated with a word cloud that
represents text from the document that relates to the concept, providing
a contextual overview.
Through a set of multiple coordinated views of text, structural
overviews, and keyword-in-context (KWIC) views, \name\ helps users
navigate a document from a specific domain perspective.
\name\ can also be used to visually and conceptually compare multiple
documents using the same domain ontology as a reference.  To aid a
domain novice, we also provide the user with navigable tooltips that
provide concept explanations that link to external references.

We illustrate the utility of \name\ by building a prototype
application\footnote{The source code of our prototype system is
available at \url{https://github.com/Xiaoyu1993/ConceptScope/}} that
visualizes computer science-related documents such as research abstracts
and articles using the Computer Science Ontology (CSO) as its reference.
Through a set of use-case scenarios, we highlight the navigation,
exploration, and comparison functions afforded by the technique, and
discuss its extension to other domains and scenarios.
We also present a brief comparison of \name\ with
DocuBurst~\cite{Collins2009docuburst} through a qualitative,
between-subjects study.
Based on our observations, we find that \name's ontology-based
visualization and its grouping of concept-related word clouds in the
Bubble Treemap helps participants define and contextualize concepts, and
explore new concepts related to a given concept.
However, \name's domain-dependency makes it less suitable for viewing
and comparing documents that span domains.
\vspace{-2.5mm}

\section{Related Work}
This paper proposes an interactive knowledge-based overview
representation of text content.
For our approach, we draw from existing techniques to identify themes or
topics in the text, and visual representations of these topics.
In this section, we outline existing work in this area and explain our
reasoning behind our choice of inspiration from the existing work.
\vspace{-2.5mm}

\subsection{Thematic Visualizations of Document Content}
Initial approaches to providing overview visualizations of document
content used metrics such as sentence length, Simpson's Index, and
\emph{Hapax Legomena} as ``literature fingerprints'' to characterize
documents~\cite{Keim2007literature}.
This approach was later used to create a visual analysis tool called
VisRA~\cite{Oelke2011visual} that helped writers review and edit their
work for better readability using these representations.
Among less abstract representations,
Wordle~\cite{Viegas2009participatory} is the most popular.
Wordle represents a text corpus as a cluster of words called a
\emph{word cloud}, with each word scaled according to its frequency of
occurrence in the text.
This idea is adapted to other techniques to characterize document
content and structures within text, such as the Word
Tree~\cite{Wattenberg2008wordtree}, which aggregated similar phrases in
sentences in a text, Phrase Nets~\cite{Van2009mapping} that visualized
text as a graph of concepts linked by relationships of the same type
found in the text, and Parallel Tag Clouds~\cite{Collins2009parallel},
that show tag clouds on parallel axes to compare multiple documents.
\vspace{-1.0mm}

When examining multiple text documents, it is important to identify the
various types of connections between them.
One of the most well-known tools used to identify inter-document
connections is Jigsaw~\cite{Stasko2008jigsaw}, which uses names,
locations, and dates to show list, calendar, and thumbnail views of
multiple documents.
While Jigsaw simply uses text occurrences to form the connections, more
sophisticated approaches have since been proposed.
Tiara~\cite{Wei2010tiara}---another system designed for intelligence
analysis---uses topic modeling with a temporal component to highlight
the change in document themes over time.
ThemeDelta~\cite{Gad2015themedelta} allows thematic comparison between
multiple documents (or similar documents over time) by combining word
clouds with parallel axis visualizations.

More recently, topic modeling-based approaches have been incorporated to
provide thematic overviews of text content.
For instance, TopicNets~\cite{Gretarsson2012topicnets} uses a
graph-based representation where both documents and topics are nodes and
links exist between documents and topics, thus serving to form clusters
of thematically-related documents.
Serendip~\cite{Alexander2014serendip} refines this idea and provides a
multi-scale view of text corpora.
It uses topic modeling along with document metadata to view patterns at
the corpus level, text level, and word level.
Oelke et al.~\cite{Oelke2014comparative} use a topic model-based
approach to compare document collections, using what they call a
``DiTop-View'' with topic glyphs arranged on a 2D space to represent the
document distribution.
ConToVi~\cite{El2016contovi} is a more recent work that uses topic
modeling on multi-party conversations to reveal speech patterns of
individual speakers and trends in conversations.
While topic model-based approaches are useful for identifying themes
within collections of documents, a knowledge-based approach requires the
use of human-organized representations of information, which are
discussed in the following section.
\vspace{-2.5mm}

\subsection{Knowledge-Based Visualizations}

As structured knowledge representation
models~\cite{Glueck2015phenoblocks}, ontologies are widely used in the
field of medicine/biology~\cite{Glueck2015phenoblocks},
engineering~\cite{Roh2016ontology, Witherell2007ontologies},
sociology~\cite{Henry2007nodetrix}, computer
science~\cite{Storey2001jambalaya}, and so on.
Achich et al.\ ~\cite{Achich2017ontology} review different application
domains and generic visualization pipelines of ontology visualization.

According to various application fields and utilizing purpose, there are
multiple methods to visualize the knowledge stored in an ontology.
The review of Katifori and Akrivi~\cite{Katifori2007ontology}
systematically categorized these methods according to the dimension of
the visualization.
Ten years later, Dud\'{a}\v{s} et al.~\cite{Dudas2018ontology} further
extended this work by adding more recently emerged visualizations.
Among these visual encodings, we find inspiration in the matrix view of
NodeTrix~\cite{Henry2007nodetrix}, the sunburst view of
PhenoBlocks~\cite{Glueck2015phenoblocks}, and the context view of
NEREx~\cite{El2017nerex}.

Our work is inspired by DocuBurst~\cite{Collins2009docuburst}, which was
the first visualization from the point of view of a human-organized
structure of knowledge.
DocuBurst uses hyponymy, or ``is-A'' relationship in the English lexicon
to identify hierarchical relationships within a given document, or when
comparing two documents.
The hierarchy is visualized as a sunburst diagram supported by
coordinated views of text content and keyword-in-context views.
While DocuBurst uses WordNet---a lexical database of the English
language---as its reference, we use domain ontologies as ours, in order
to provide a more focused, domain-specific overview of documents.
\vspace{-2.5mm}

\subsection{Hierarchical Layouts}

Visualization of a knowledge-based document overview needs to
incorporate the hierarchical information inherent to the knowledge base.
While a tree is the common representation of such a hierarchy, it is
usually more suitable for showing the structure rather than the content
of the information presented.
The most famous alternative for representing hierarchical information is
the TreeMap~\cite{Shneiderman1992tree}, a two-dimensional, space-filling
layout that represents hierarchy through nesting and a second quantity
such as percentage contribution to the whole as the area.
Alternatives to TreeMaps such as Icicle plots and Radial
TreeMaps~\cite{Barlow2001comparison} and Sunburst
diagrams~\cite{Stasko2000evaluation} have since been proposed and
incorporated into standard visualizations of hierarchies.
DocuBurst~\cite{Collins2009docuburst} referenced in the previous section
uses the Sunburst diagram as its hierarchical visualization.

While the original TreeMap has afforded enough space in the
representation to portray content, it often comes at the cost of some
loss of detail in the hierarchy.
Alternatives such as circle packing~\cite{Wang2006visualization} and
more recently, Bubble Treemaps~\cite{Gortler2017bubble} have been
proposed to address this issue.
We incorporate the Bubble Treemap into our design for its relative
compactness compared to circle packing, and its use of space that allows
for some content representation.

\section{Requirements and Design}
In this section, we break down our overall need to provide a
knowledge-based overview of document content into specific requirements
to inform the design of \name.
We apply Collins et al.'s~\cite{Collins2009parallel} question ``What is
this document about?'' to the general ``distant reading'' tasks for both
single text analysis and parallel text analysis posed in J{\"a}nicke et
al.~\cite{Janicke2015close}, typically addressed using intrinsic text
properties such as entity/location occurrences, text frequencies, etc.,
but not domain knowledge. 
Our requirements stem from exploring tasks of hierarchical overview,
document comparison, and concept exploration using a knowledge base as
reference.

\begin{enumerate}[itemindent=0.65em, itemsep=1.5mm]
  \item[\textbf{R1}] 
    \textbf{Provide Conceptual Overview:}
    When reading a long document from an unfamiliar domain---such as an
    academic paper---the reader can benefit from a high-level overview
    of the information provided.
    While word clouds can provide a simple overview of the \emph{text}
    in the document, a lack of understanding of the technical terms
    might hinder the reader in understanding the overview
    representation.
    Instead, an overview that stems from a fundamental categorization of
    the domain itself---as represented by the hierarchical organization
    of concepts often available in an ontology---can provide an overview
    that is accessible to both novices and experts in the domain.

  \item[\textbf{R2}] 
    \textbf{Reveal Contextual Information:}
    The document text and the ontology do not always overlap.
    From the point of view of the ontology, the document contains
    non-relevant information, but information nevertheless important for
    the reader.
    For instance, a research paper introducing a new search algorithm
    can introduce several concepts in the knowledge base of search
    algorithms.
    The paper would also make arguments for and against certain
    algorithms.
    The reader may benefit considerably from the structure and content
    of these arguments, which are lost if the overview visualization
    focuses solely on the ontological components.
    A way to provide the contextual information surrounding these
    concepts is thus needed.

  \item[\textbf{R3}] 
    \textbf{Support Exploration of New Knowledge:}
    When exploring a concept that is a sub-domain of a domain that is
    only partially known to the reader, they may be interested in other
    sub-domains of the domain.
    For example, if the term ``quicksort'' appears in an algorithm
    paper, the reader might want to know of other sorting algorithms
    such as ``bubble sort'' and ``merge sort''.
    They may also want to learn about related terms such as ``divide and
    conquer'' and ``time complexity''.
    These new terms may not appear in the document text, but forms an
    essential component of knowledge that extends from---and aids the
    understanding of---the core concept (i.e.\ quicksort).
    We thus need ways to enable users to access information from the
    ontology that is related to the concept of interest.
    
  \item[\textbf{R4}] 
    \textbf{Support Multi-document Comparison:}
    Document comparison is a common requirement that emerges from the
    creation of visual overviews of
    documents~\cite{Collins2009docuburst}.
    In the case of our scenario, the comparison is likely to be
    conceptual: to get a quick comparison of concepts that are common to
    multiple documents, and those that are unique to one.
    The reader may also want to simply compare the differences between
    the information provided in two documents.
    While documents such as academic papers may contain an abstract that
    summarizes the main content of the article, it may not be sufficient
    enough to cover all the concepts that are covered in the papers, not
    to mention the similarities and differences.
    Therefore, our tool should be able to provide visual support for
    users to compare and analyze the conceptual structure and content
    between two or more documents.

\end{enumerate}
\vspace{-2.5mm}

\section{Implementation}
\begin{figure*}[tb]
  \centering 
  \includegraphics[width=0.8\textwidth]{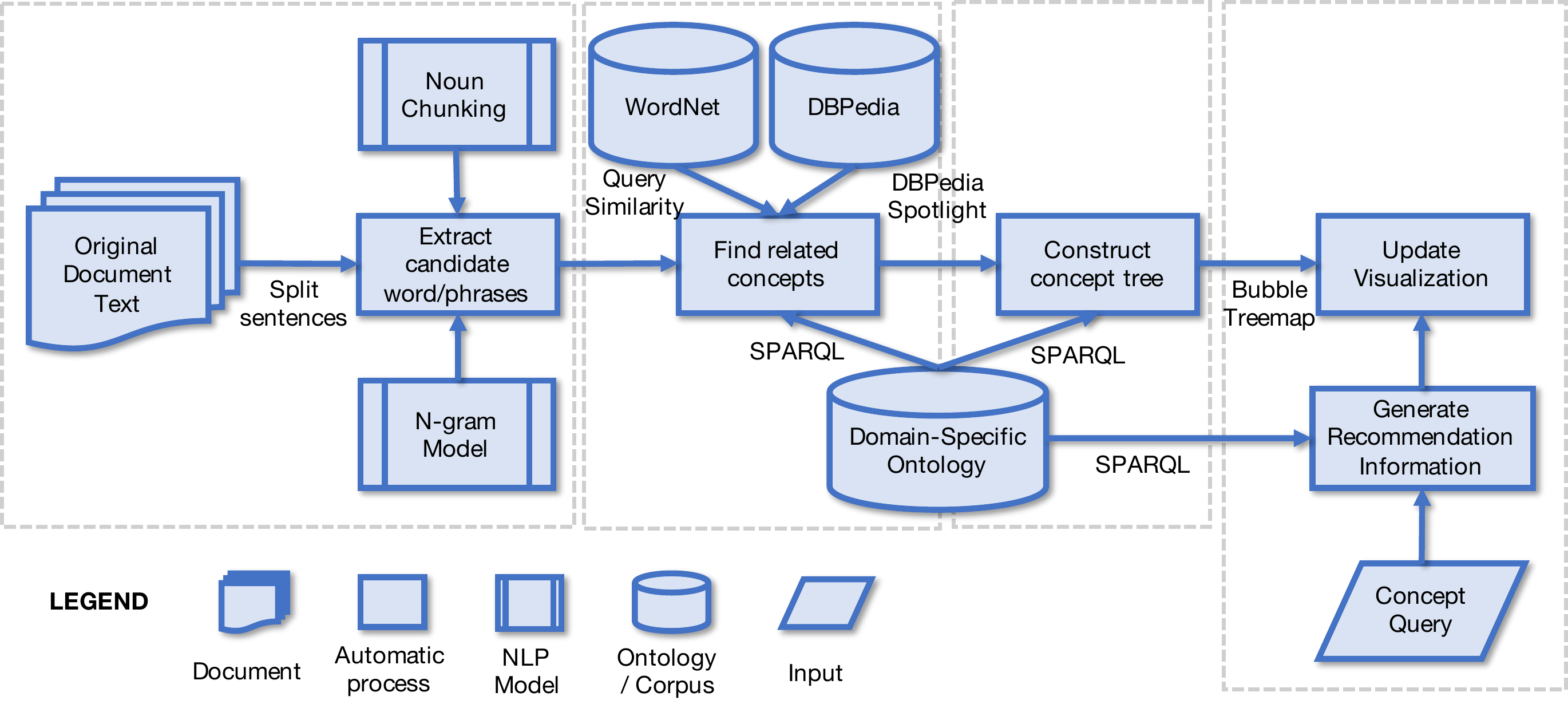}
  \caption{Data processing pipeline for \name.}
  \vspace{-5mm}
  \label{fig:pipeline}
\end{figure*}

In order to provide the knowledge-based conceptual overviews of a given
document, an appropriate mechanism is needed to parse the document and
compose queries to the reference ontology.
An appropriate representation of the concept needs to be automatically
generated in a way that reflects its hierarchy in the domain ontology as
well as its occurrence in the document.
To achieve this, we need to incorporate techniques from multiple areas
including natural language processing, ontology querying, and
information visualization.
Figure~\ref{fig:pipeline} shows the framework of assembling them into a
pipeline and the section number describing the corresponding technical
details.
\vspace{-2.5mm}

\subsection{Generating Query Candidates}
\label{sec:candsearch}

Ontology queries are typically performed using SPARQL (SPARQL Protocol
And RDF Query Language)~\cite{Sparql}, which typically use ``triples''
(subject, predicate, and object) or parts thereof.
In our case, trials showed that an exact triple was unlikely to be
constructed from the document, nor was it deemed necessary. 
Instead, it was more important to have the subjects or objects be
specific terms that are likely to be present in the ontology.
We construct these queries from the document with a sentence-level
granularity.
In order to construct the query terms, we use two approaches: noun
chunking, and n-gram identification.

Noun chunking is the process of extracting subsets of noun phrases such
that they do not contain other noun phrases within
them~\cite{Bird2009natural}.
This allows us to identify specific terms that may be relevant to a
domain ontology.
For instance, when referencing the computer science ontology, terms such
as ``object-oriented programming'' and ``local area network'' are much
more meaningful than the individual words that make up these terms
(``local'', ``object'', or ``area'').
For this reason, we also do not resort to stemming or lemmatization as
they change the morphology of the word (e.g., ``oriented'', if
lemmatized to ``orient'', forms ``object-orient programming'') which
renders the noun chunk invalid as a query candidate.
Noun chunks can also include leading or trailing stop words, which are
trimmed in order to generate the query candidates.

Noun chunking can produce phrases that contain query candidates but are
not query candidates themselves.
For instance, a paper about animation may include multiple variances of
animation like ``2D computer animation'', ``stop-motion animation'' and
``animated transition''.
Some of these may appear within noun chunks, but not by themselves.
To identify such cases, we identify groups of words that commonly occur
together in the document as n-grams.
\vspace{-2.5mm}

\subsection{Mapping Queries to Concepts}
\label{sec:conceptmap}

Once the query candidates are identified, the next step is to map these
candidates to the corresponding concepts in the domain ontology of
interest.
This involves two steps: (1) perform identical matches, i.e.\
concepts that correspond exactly to those in the ontology, and (2)
reduce the number of ``failed'' matches, i.e.\ concepts that are related
but not present in the ontology.
Step 2 is often necessary as domain ontologies are not all uniformly
mature.
For instance, Computer Science Ontology is not as well-populated as,
say, medical or biological ontologies such as the human phenotype
ontology.

The two steps---accurate matching and fuzzy matching---are illustrated
in lines $8$ through $15$ in Algorithm~\ref{alg:conceptdetection}.
For any given candidate, we first look for an accurate match in the
domain-specific ontology.
We then construct a dictionary that includes all of the concepts in the
ontology for an effective search.
However, the number of concepts that can be directly detected by
accurate matching is small.
This is because of the mismatch between specific forms in which a
concept is listed in the ontology and its many variations in the
document.
For instance, ``object-oriented programming'' may be the exact match in
the ontology, but it might appear in the text as ``object-oriented
approach'' which is clearly related but cannot be identified with an
accurate match.
In order to solve this problem, we introduce a fuzzy match.
\vspace{-3mm}

\begin{algorithm}
 \caption{\textbf{Detect CSO Concepts in Document}}
 \begin{algorithmic}[1]
 \renewcommand{\algorithmicrequire}{\textbf{Input:}}
 \renewcommand{\algorithmicensure}{\textbf{Output:}}
 \REQUIRE  document text $stringDoc$
 \ENSURE  concept dictionary $dictConcept$
 \STATE $listSent \leftarrow Split\left( stringDoc\right)$
 \STATE $modelNGram \leftarrow TrainNGram\left(listSent\right)$
 \STATE $dictConcept \leftarrow \emptyset$
 \STATE \textbf{for} $stringSent$ \textbf{in} $listSent$: \algcom{// iterate over each sentence of the document}
 \STATE \ \ \ \ $listNGram \leftarrow modelNGram\left(stringSent\right)$ \algcom{// identify initial query terms}
 \STATE \ \ \ \ $listChunk \leftarrow NounChunking\left(stringSent\right)$ \algcom{// identify additional query terms 
 }
 \STATE \ \ \ \ $listCand \leftarrow \  listChunk \ \cup\ listNGram$
 \STATE \ \ \ \ \textbf{for} $stringCand$ \textbf{in} $listCand$: \algcom{// iterate over each candidate query term}
 \STATE \ \ \ \ \ \ \ \ \textbf{if} $QueryCSO\left(stringCand\right) \neq \emptyset$: \algcom{// accurate matching}
 \STATE \ \ \ \ \ \ \ \ \ \ \ $dictConcept \leftarrow dictConcept \cup QueryCSO\left(stringCand\right)$
 \STATE \ \ \ \ \ \ \ \ \textbf{else}: \algcom{// fuzzy matching}
 \STATE \ \ \ \ \ \ \ \ \ \ \ $fuzzyCand \leftarrow DBpediaSpotlight\left(stringCand\right)$ \algcom{// get candidate DBpedia concepts}
 \STATE \ \ \ \ \ \ \ \ \ \ \ $fuzzyCand \leftarrow Filter\left(fuzzyCand,\  threshold\right)$ \algcom{// filter candidate DBpedia concepts according to similarity}
 \STATE \ \ \ \ \ \ \ \ \ \ \ \textbf{if} $QueryCSO\left(fuzzyCand\right) \neq \emptyset$: \algcom{// link the filtered DBpedia concepts back to CSO concepts}
 \STATE \ \ \ \ \ \ \ \ \ \ \ \ \ \ \ \ $dictConcept \leftarrow dictConcept \cup QueryCSO\left(fuzzyCand\right)$
 \end{algorithmic}
 \label{alg:conceptdetection}
 \end{algorithm}
\vspace{-3mm}

The goal of fuzzy matching is to match the candidate to a concept
that is very close to but not exactly equal to the candidate.
In our prototype system, we use the computer science ontology (CSO) as
the domain-specific ontology.
The CSO also incorporates links of the form \emph{``sameAs''}
(\href{http://www.w3.org/2002/07/owl\#sameAs}{http://www.w3.org/2002/07/owl\#sameAs}),
that connect to DBPedia~\cite{Lehmann2015dbpedia}, a broader, but less
strictly-defined and less domain-specific ontology.
We use these links and leverage the DBpedia Lookup
Service~\cite{DBpedialookup} to find related DBpedia concepts and link
them back to CSO.
After checking the semantic similarity between the CSO concept detected
in this way and the original candidate query term using the
Wu-Palmer similarity measure offered as a default function in
WordNet~\cite{Miller1998wordnet}, we add the concept to the dictionary
if that similarity is above a threshold.
This threshold is currently determined by trial and error.
\vspace{-3mm}

\subsection{Hierarchy Reconstruction}
\label{sec:dataconstruct}

The concept dictionary constructed thus far does not yet incorporate
hierarchical information.
In order to retrieve and store the hierarchical information from
the ontology, we query the paths from every detected concept to the root
of the ontology and use them to restructure the concept dictionary as a
tree.
The final output of this algorithm---the concept tree---can be directly
converted to a JSON file and used to automatically render the
visualization.
\vspace{-1mm}

\section{\name\ Interface}
In this section, we discuss the visualization design and the
interactions supported in \name.
\vspace{-3mm}

\subsection{Visual Encoding}
\label{sec:vsl_encdg}

We choose Bubble Treemaps proposed by G{\"o}rtler et
al.~\cite{Gortler2017bubble} as our primary visualization.
This visualization is originally designed for uncertainty visualization,
but we find it suitable for our application in terms of hierarchy
representation and space organization (\textbf{R1}).
We use the original layout algorithm of the Bubble Treemap, but adapt
the visual encoding and interaction strategies to meet our design
requirements.
\vspace{-2mm}

\subsubsection{Hierarchy Presentation}
\label{sec:hierarchy}

In a Bubble Treemap, the deepest levels of the hierarchy are represented
as circles, with successively higher levels forming contours around
their ``child'' levels.
We use the circles to represent the terms that appear (or have
corresponding synonyms) in the original document as well as in the
ontology.
The outer contours represent concepts that do not explicitly appear in
the document but still represent parent concepts from the ontology.
These parent concepts are identified using the ontology query process
demonstrated in Algorithm~\ref{alg:conceptdetection}.
The outermost contour forms the ``root'' of the ontology, with
successive inner contours representing its child concepts.
For example, in the computer science ontology
(CSO)~\cite{Salatino2018computer} we use for our case studies, the term
``computer science'' is the root concept in the ontology. 
\vspace{-2mm}

\begin{figure*}[t]
  \centering
  \includegraphics[width=1.0\textwidth]{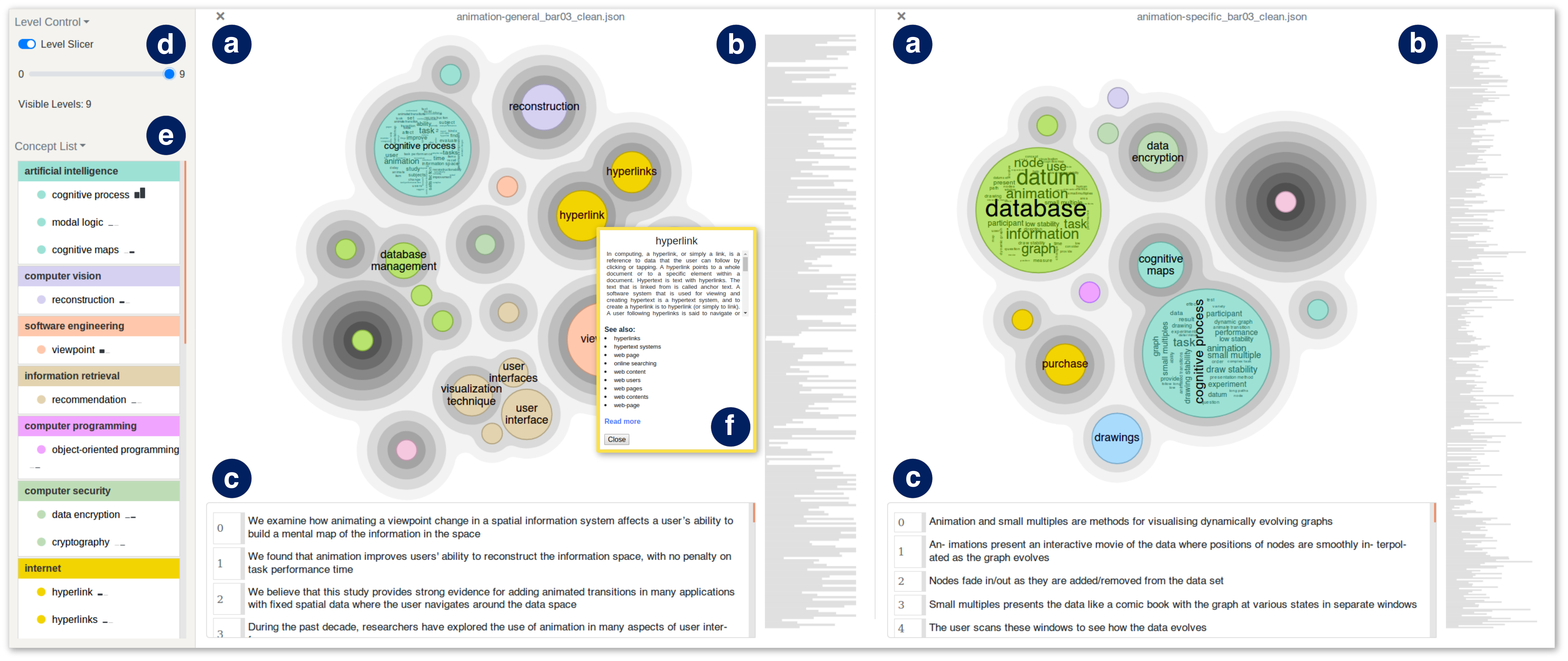}
  \caption{
    The \name\ interface representing two research papers discussing
    animation.
    The Bubble Treemaps (a) provide overviews, with the one on right
    showing a paper covering more specific topics than the one on the
    left.
    Supporting transcript (b) and text (c) views, along with a level
    slicer (d) and a list presentation (e) allow exploration and
    comparison between the documents, while a tooltip (f) allows
    examination of a concept of interest.
 }
    \vspace{-5mm}
 \label{fig:teaser}
\end{figure*}

\paragraph{Inner Circles}
The function of the innermost circles---representing concepts that are
present in the ontology \emph{and} in the document---is to provide a
clear representation of the terms that are directly connected to the
document.
The size of the circles is proportional to the frequency with which the
corresponding term appears in the document. 
The fill color of a given circle corresponds to the highest ``parent
concept'' it belongs to, just below the root.
Although the Bubble Treemap layout already gathers together circles that
share the same parents, we visually reinforce such relationships by
assigning the same color to circles with the highest common ancestor
(besides the root).
These ``highest parent concepts'', divide the root term into several
subclasses and help users to better grasp the various areas the document
covers.
In order to make sure the circles' colors are perceptually uniform, we
create the isoluminant palette~\cite{Kovesi2015good} from the CIELAB
color space to ensure perceptual uniformity between the concepts shown.
\vspace{-2mm}

\paragraph{Surrounding Contours}
The contours surrounding the circles show hierarchical relationships
between the concepts that occur in the document.
After exploring several encoding options for the contours to best
represent related concepts while highlighting hierarchies, we chose fill
colors of decreasing luminance to represent ``deeper'' contours in the
hierarchy.

\subsubsection{List Presentation}
\label{sec:scent_leg}

Effective as the Bubble Treemap is, it is not intuitive enough for the
users to understand and grasp all necessary information at a glance
(\textbf{R1}).
Therefore, we augment the visualization with a multi-function widget
(Fig.~\ref{fig:teaser} (e)) which combines concept list, legend,
and bar charts representing term frequencies to solve this problem.
Inspired by scented widgets~\cite{Willett2007scented}, the
multi-function widget presents important supporting information in a
compact representation.
As a concept list, this tool represents every concept detected in the
currently-loaded document(s) as a list item, the background color of
which is the same as the corresponding concept circle(s) shown in the
Bubble Treemap.
We group the concepts sharing the ``highest super topic'' together, with
an additional list item showing the common ``highest super topic'' of
each group.
This concept list also acts as a legend showing the connection between
each color and their corresponding ``highest super topic''.
We also attach a sparkline for each list item to show the distribution
of current concept across multiple documents (when multiple documents
are loaded).

\subsubsection{Incorporating Word Clouds}
\label{sec:word_clouds}
An unlabeled Bubble Treemap can be too abstract a representation for the
user to comprehend.
On the other hand, labeling every concept may result in a cluttered view
which would also make comprehension difficult.
We thus provide three levels of labeling for the concept: unlabeled (if
the concept circle is too small), labeled (if the concept circle is
large enough to fit its corresponding concept name), and labeled with
context (where a word cloud of related terms from the document is
combined with the concept label) (\textbf{R2}).
The interactions to control these views are discussed in the following
section.

\subsection{Interaction}
\label{sec:interaction}

\name\ provides linking between views and semantic overview and detail
views to help analyze the document(s) and its concepts.
These interactions support two modes of document analysis: exploration
and comparison.
We will first describe the overview and detail interactions and follow
them with the modes of analysis.

\subsubsection{Overview+Detail Interactions}
\label{sec:int_sz}

To eliminate the potential confusion caused by the users' unfamiliarity
with the Bubble Treemap, we introduce interactions to acquaint them with
the visual schema and provide details on
demand~\cite{Shneiderman2003eyes}.
The Bubble Treemap provides a compact view of the domain-relevant
concepts, their hierarchical structure in the ontology, as well as their
context in the original document.
In order to make this compact representation easier to understand, we
design two interactions to present information that the user may seek:
(1) a \emph{level slicer} to ``slice'' the Bubble Treemap at any level
to examine parent concepts, and 
(2) \emph{semantic zooming}, which allows the user to zoom in to a
concept circle to examine its corresponding word cloud (described in
Sec.~\ref{sec:word_clouds}).
The users can choose and combine these two tools according to their
preference.

The \textbf{Level Slicer} is designed to help novice users quickly build
a connection between the nested layout of the Bubble Treemap and the
hierarchical structure of ontology (\textbf{R2, R3}).
This tool allows the user to choose the level of the parent concept that
they want to see on the screen by sliding the slider bar.
When the view initializes, all levels of the Bubble Treemap are shown to
provide an overview, but the labels corresponding to parent contours are
concealed.
Once the ``child'' concepts are sliced away by the slicer, the
corresponding labels of the newly exposed parent concepts are made
visible.
This tool facilitates users to inspect any cross section from the whole
hierarchical structure that interests them.

\textbf{Semantic Zooming} is designed to provide different granularities
of information based on users' needs (\textbf{R2, R3}).
As explained in Sec.~\ref{sec:word_clouds}, users may see three levels
of detail for the same concept circle: \textit{unlabeled},
\textit{labeled}, and \textit{labeled with word cloud}.
When users zoom in and out of the graph, the size of every circle
changes and its appearance transforms among the three based on the
available space inside it.

\name\ also reveals more information about a concept including its
thumbnail, definition, related concepts, and its context in the
text.
These views allow the exploration of concepts that do not
themselves occur in the document but are related to the ones that do
occur (\textbf{R3}).

\subsubsection{Exploration Mode}
\label{sec:int_exp}

The exploration mode---meant for inspecting a single document---provides
conceptual overview+detail representations of the document using the
ontology as a reference.
With the static Bubble Treemap, it is almost impossible for novice users
to build the connection between a circle in the graph and a word/phrase
in the original text.
Users might want to explore related knowledge in the domain ontology
about the concepts shown in the Bubble Treemap.
Following the information-seeking mantra~\cite{Shneiderman2003eyes}, we
design a set of small widgets that can be easily evoked and interacted
with to the Bubble Treemap.

To \textit{connect the Bubble Treemap and the original document}
(\textbf{R2}), we create a high-level transcript view and a raw
text view.
The high-level transcript view can be seen as a ``minimap'' of the
document, with each sentence represented by a series of horizontal lines
scaled to sentence lengths (Fig.~\ref{fig:teaser} (b)).
In the raw text view, the raw text is shown to provide a convenient
context acquisition (Fig.~\ref{fig:teaser} (c)).
These two views as well as the Bubble Treemap view are fully
coordinated, so that interacting with one view highlights related
information in the other views.
For example, if the users hover over a circle representing a concept in
the Bubble Treemap view, the lines corresponding to the sentences that
contain this concept in the transcript view and the text of the sentence
in the raw text view are also be highlighted.

Interacting with a concept circle also reveals a tooltip that shows the
concept definition,  a link to the relevant concept page on DBPedia
(\textbf{R3})(Fig.~\ref{fig:teaser}(f)), and a thumbnail (if available
in the corresponding DBPedia entry).
The tooltip also provides links to other related concepts that may not
be present in the document, to provide context from an ontology point of
view.

\subsubsection{Comparative Mode}
\label{sec:int_cmp}

The comparative mode assists users in comparing multiple documents and
explore conceptual similarities and differences between the documents
(\textbf{R4}).
As the name suggests, loading multiple documents creates multiple,
side-by-side Bubble Treemap views, one for each document.
Concepts common to two or more documents are encoded in the same color
across the Bubble Treemaps.

The comparative mode provides similar interactions as the exploration
mode.
In addition, the sparklines mentioned in Sec.~\ref{sec:scent_leg} can
provide users with a quick overview of the relative frequency with which
each concept occurs across the documents.
The users can compare the concepts that interest them by hovering or
searching.
If they know where a concept is located in any of the Bubble Treemaps,
the user can simply hover on the corresponding circle or contour, which
highlights the concept---if available---across all the Bubble Treemaps.
They can also directly search for the concept in the search field (top
right corner in Fig.~\ref{fig:teaser}) to highlight all relevant circles
and contours across the Bubble Treemaps.
The users can thus quickly get an idea about where and how their
concepts of interest are distributed across different documents.

The switchover between exploration mode and comparative mode does not
require explicit user operation.
Loading a single document shows the exploration mode while loading
additional documents sets \name\ to comparison mode.
The exploratory features are always available regardless of the number
of documents, as comparison also requires a degree of exploration.
We also provide a ``switch'' alongside the level slicer in
Fig.~\ref{fig:teaser}(d) for semantic zooming to make sure the users can
explore or compare the Bubble Treemap(s) at whatever number of levels
and size they want.

\section{Use-Case Scenarios}
We briefly illustrate the use of \name\ for exploring and comparing documents with two use-case scenarios: exploring an academic paper and comparing the transcripts of three TED talks.
\vspace{-2.5mm}

\subsection{Exploring an Academic Paper}
\label{sec:case_exp}

\begin{figure*}[tb]
  \centering
  \includegraphics[width=\textwidth]{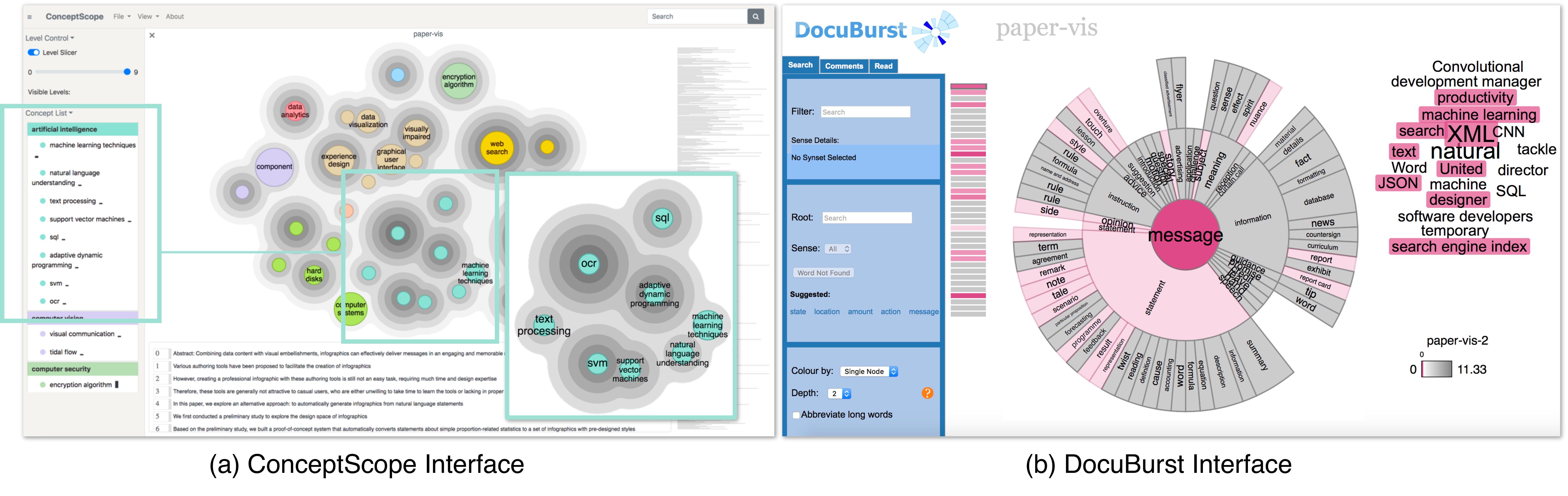}
  \caption{Overview of an IEEE VIS 2019 Paper~\cite{Cui2019text} using
    \name\ (left) and using DocuBurst (right).
  }
  \label{fig:paper_vis}
\end{figure*}

We first use \name\ to visualize an academic paper~\cite{Cui2019text} on
automatic infographics generation, published in IEEE VIS 2019.
To ensure the accuracy of our natural language processing components, we
only keep the natural-language parts of the original paper and remove
text in references, tables, formulae, and figure labels. 
We use the computer science ontology (CSO) as the reference ontology for this paper.
Fig.~\ref{fig:paper_vis} shows the visualization, with the same paper
shown in DocuBurst~\cite{Collins2009docuburst} for reference.

The Bubble Treemap shows over 30 computer science concepts 
directly or indirectly mentioned in the paper (requirement
\textbf{R1}).
Inspecting the concept list on the left, we see that the highest
parent concepts of the ones identified in the document range from 
``human-computer interaction'' to ``artificial intelligence'' to
``computer system''.
Zooming in, we click on the bubble representing ``OCR'' and a tooltip pops up with
the definition of this concept as well as the recommendation of
concepts related to this one (\textbf{R3}).
We examine the definitions and where the concept appears in the word
cloud to see that it points to the use of OCR to identify key text in
existing infographics (\textbf{R2}).
We also see that these and most concepts under ``artificial
intelligence'' appear under the related work section.
We thus infer that these concepts might only be mentioned as
background or references to other work, and not as a fundamental
contribution of the paper.

Figure \ref{fig:paper_vis} (right) shows the DocuBurst visualization
using the root ``message''.
We notice that almost all computer-science-related concepts identified
by DocuBurst can be detected by \name\ as well.
In terms of space efficiency, DocuBurst has the advantage of providing a
more compact visualization with its Sunburst diagram.
However, DocuBurst offers fewer options for contextual views.  In \name,
the word clouds in each concept circle provide a contextual overview and
aid concept exploration outside the realm of the document with our
detail-info tooltip of concepts and the links to DBPedia.
\vspace{-2.5mm}

\subsection{Comparing Transcripts of TED Talks}
\label{sec:case_cmp}

\begin{figure*}[t]
  \centering
  \includegraphics[width=\textwidth]{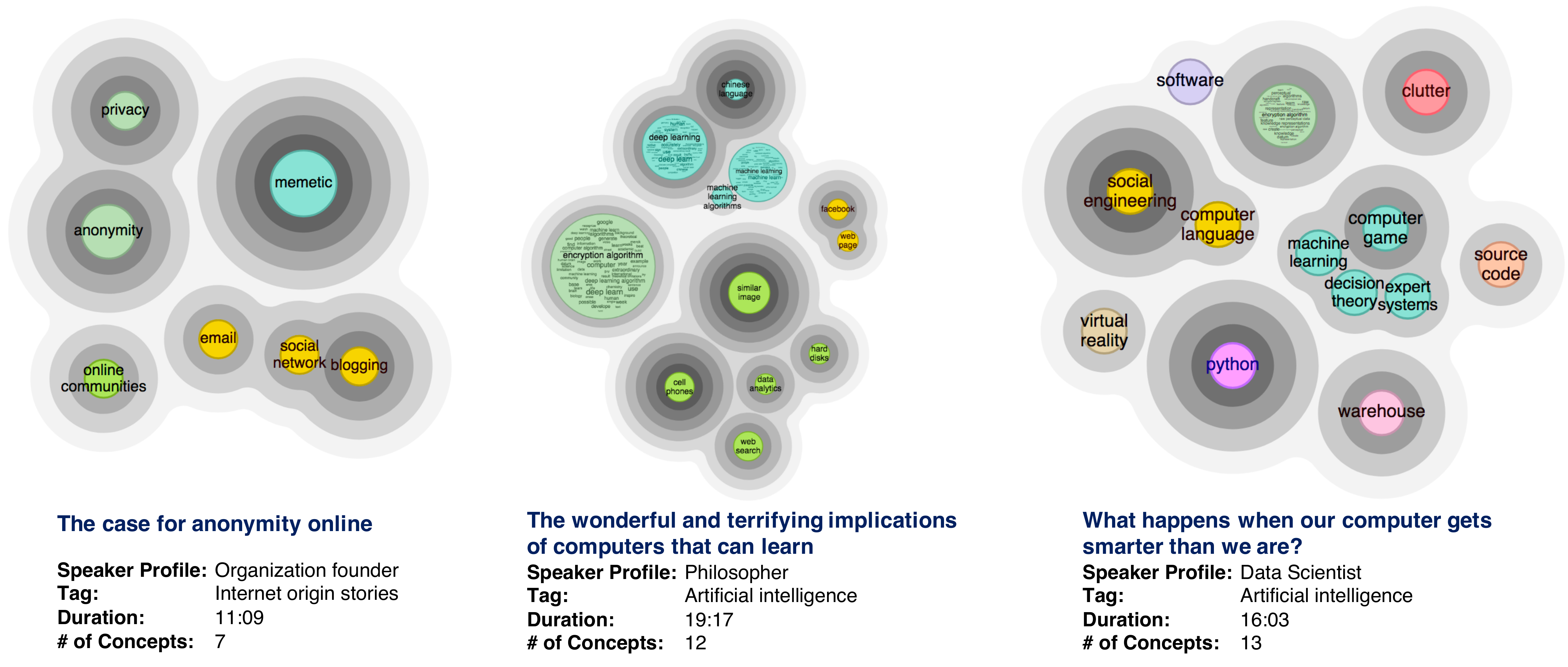}
  \vspace{-7mm}
  \caption{\name\ visualizations comparing the transcripts of three TED
    Talks.
    The title of each talk is shown in red under each visualization,
    along with speaker profile and talk metadata.
    Also shown is the number of concepts from CSO found in each
    document.
  }
  \label{fig:ted_talks}
  \vspace{-5mm}
\end{figure*}

To illustrate multi-document comparison, we load the transcripts of
three TED Talks~\cite{Poole2010case, Bostrom2015what,
Howard2014wonderful}, all of which are tagged under the ``computers''
category on the TED webpage.
Fig.~\ref{fig:ted_talks} shows the distribution and depth of concepts,
along with information about each talk.

Loading all three documents into \name\ creates three panels (similar to
that shown for two papers in Fig.~\ref{fig:teaser}), each containing the
Bubble Treemap view, transcript view, and raw text view for the
corresponding transcript.
The Bubble Treemap immediately illustrates the differences and
similarities between concepts across the three talks, which can further
be explored as all three views are coordinated.
We notice that all three of the talks mention concepts under the parent
topics of ``internet'', ``computer security'' and ``artificial
intelligence''.
One reasonable explanation is that these topics cover many basic terms
in computer science, so it is almost unavoidable to use them in a
computer-science-related technical presentation.
When inspecting the concept list and Bubble Treemaps, we notice that
concepts that belong to ``artificial intelligence'' appear more in talk
No.\ 2 and talk No.\ 3, which makes sense as the two talks have the
additional tag of ``AI'' on the TED webpage. 

Talk No.\ 1 discusses the issue of privacy on online forums, and
concepts of privacy and anonymity fall outside the current version of
the computer science ontology.
In addition, the talk does not delve deep into computer science concepts.
This results in a Bubble Treemap that covers very few concepts.
Talk No.\ 2 is delivered by a data scientist who talks about computer
science concepts, specifically ``algorithms'', ``machine learning'', and
``deep learning'', which are reflected in the Bubble Treemap.
Finally, Talk No.\ 3 is presented by a philosopher who talks about
broader implications of machine learning, also providing a historical
perspective.
This is reflected in the Bubble Treemap, showing the broadest concept
coverage of the three talks, with no one concept being too dominant.

\section{Study}
We conducted a controlled study to evaluate whether the visualization \&
interaction design and the use of a domain-specific reference ontology
renders \name\ effective in exploring single documents or comparing
multiple documents.
Specifically, we intended to understand whether \name\ was effective in
helping users:
(1) summarize the content of a document 
with a domain-specific concept overview (\textbf{R1}); 
(2) glean what a document says about any given concept in the context of
the document (\textbf{R2});
(3) become aware of new concepts and their connections (\textbf{R3});
and
(4) discover enough similarities and differences among multiple
documents (\textbf{R4}).
In order to provide a baseline, we used
DocuBurst~\cite{Collins2009docuburst}, the popular content-oriented
document visualization tool that provides a non-domain-specific overview
of documents using the WordNet~\cite{Miller1998wordnet} taxonomy.
We thus conducted a between-subjects study comparing participants that
used \name\ with participants that used DocuBurst.
Please note that the generalizability of this study might be affected by
the limited number of participants we could recruit and the diverse
devices they used due to safety measures surrounding COVID-19. 
However, the way we report the insights were mainly based on patterns
and not numbers, so the validity is not highly impacted by those
factors.

\subsection{Participants}
\label{sec:study_par}

We recruited 18 participants (10 female, 8 male) aged between 18 and 44
years.
The participants comprised 16 Ph.D. students, 1 undergraduate student,
and 1 employee of a technology company. 
Seventeen participants had computer science backgrounds, of which 12
specialized in visualization and HCI, 1 in high-performance computing, 1
in natural language generation and multi-modal learning, while 3 didn't
report their specialized field.
The one remaining participant had a design and education background,
specializing in learning and user experience design.
Two of the 18 participants reported themselves as native English
speakers.

\subsection{Conditions and Task Design}
\label{sec:study_task}

Most document visualization systems use either intrinsic statistical
information such as topic models and word co-occurrences, or
human-curated categories that do not scale to large knowledge bases
(e.g.~\cite{Nualart2013texty}).
Per Kucher et al.'s survey~\cite{Kucher2015text}, which is currently up
to date\footnote{Text Visualization Browser:
\url{https://textvis.lnu.se/}}, DocuBurst is the only knowledge-based
document exploration system.
We thus chose DocuBurst as the baseline for our evaluation.

DocuBurst provides an overview of documents based on the non-domain-specific ``is-a'' relationship in WordNet, while our prototype is based on domain-specific ontologies, in this case, the Computer Science Ontology (CSO).
We asked each participant to perform the same tasks using the interface assigned to them (\name\ or DocuBurst) and compared interaction and behavior patterns across participants.
Participants were given time to familiarize themselves with their assigned interface.
They were then asked to perform the following tasks:

\begin{enumerate}[itemindent=0.65em, itemsep=1.5mm]

  \item[\textbf{T1}] 
    \textbf{Explore one single document:}
    This task was divided into several sub-tasks, each aligned with a
    corresponding design requirement:
    (1) summarize the documents and provide relevant keywords
    (\textbf{R1});
    (2) describe a specified concept based on its usage in the document
    (\textbf{R2});
    (3) select (from a list of description) the context in which a given
    concept is used in the document (\textbf{R2});
    (4) define several concepts before and after using the system, as
    well as rate confidence with the definition (\textbf{R3});
    (5) identify concepts in the document related to a given concept
    (\textbf{R3}); and
    (6) list the concepts (that the participants did not know before the
    study) in the document (\textbf{R3}).
    Participants were also asked whether they read the documents before
    the study to account for potential confounds.
      
  \item[\textbf{T2}] 
    \textbf{Compare two documents:}
    The participants were asked to compare two documents at a conceptual
    level (\textbf{R4}).
    Therefore, they were asked to identify common and unique concepts,
    as well as overall similarities and differences between the two
    documents.
    Again, they were asked whether they read the documents before the
    study to eliminate bias.
    
  \item[\textbf{T3}] 
    \textbf{Compare three documents:}
    The questions that participants were asked to answer in this task
    were generally the same as task T2 but for three documents.
    One difference was that participants were suggested to ``identify a
    theme and explain their difference within the theme'' when
    identifying the difference between three documents.
    Since DocuBurst was not capable of comparing more than two
    documents, this task was only assigned to participants using \name\
    in the study.
    
\end{enumerate}

In order to mirror participants' regular reading experience, we chose
computer-science related academic papers or technical reports for all
tasks of this study.
For task T1 we used Munzner's nested model for validating
visualizations~\cite{munzner2009nested}. 
Task T2 involved two papers discussing animation techniques: the first,
a general evaluation of how animation could help users build a mental
map of spatial information~\cite{Bederson1999does}, while the other
focused on the role of animation in dynamic graph
visualization~\cite{Archambault2016can}.
To alleviate participant fatigue and manage their time, we chose to use
relatively shorter transcripts of three 15--20 minute Ted
Talks~\cite{Tufekci2016machine, Bostrom2015what, Howard2014wonderful} in
the ``artificial intelligence'' playlist instead of academic papers for
task T3.

\subsection{Study Setup}
\label{sec:study_exp}

We conducted the study remotely owing to safety measures surrounding
COVID-19.
The participants were asked to access either of the tools from a remote
server and participate in the study with their own machine and external
devices.
Fourteen of them used laptops with screen sizes ranging from 13 in.\ to
16 in.
The other 5 used monitors with screen sizes ranging from 24 in.\ to 32
in.
Fifteen participants used the Chrome browser, 2 used Safari, while one
used Firefox for the tasks.

The setup, tasks, and durations were decided based on a within-subjects
pilot of the study described above with 2 participants: one native and
one non-native English speaker.
\name\ and DocuBurst employed different datasets in this study.
The decisions to suggest time durations for the questions and to set up
the final study as a between-subjects study were made based on the long
duration of each session and on the participant's fatigue toward the end
of each session.

\subsection{Procedure}
\label{sec:study_proc}

Participants first responded to an online pre-survey providing their
demographic and background information.
Once they had finished familiarizing themselves with the interface, the
participants performed the tasks described in Sec.~\ref{sec:study_task}.
Participants followed a concurrent think-aloud protocol while executing the tasks, with the moderator recording their verbalizations and their screen through a videoconferencing application.
Finally, the participants were invited to finish a brief survey about
the tool and share their feedback about their experience with the
interface, both as open-ended responses and on the NASA TLX
scale~\cite{Hart1988development}.

\section{Results and Discussion}
\subsection{General Behavior Patterns}
\label{sec:result_bhv }

We categorized participants into two groups based on how they attempted
to gather the information they needed to answer the questions, rather
than how they used the tools in general.
One group comprised participants that mainly used the visualization, and
the other, those that mainly used the raw text display.
Seven of the 9 participants who used \name\ primarily used the main
Bubble Treemap visualization to glean the required information, while
the remaining 2 relied more on the raw text reading from the document.
In DocuBurst, only 5 of the 9 participants used the main Sunburst
diagram as their main source of information, while 4 chiefly relied on
close-reading of the text.

Participants using \name\ used the main visualization more than
participants using DocuBurst.
This was partly due to the raw text reading experience offered by the
two interfaces, and partly due to the ability of the visualizations and
the knowledge base in conveying a relevant overview.
In \name, documents were split into sentences  and displayed in a
relatively small vertical space (see Fig.~\ref{fig:teaser}c).
Therefore, participants tended to read only a few sentences prior to and
after the key sentence for a specific task instead of going through
larger blocks of text.
As participant $Pc7$ stated, \textit{``because my resolution is small
and my mouse is sensitive, so when I move it jumps between the text very
easily (in transcript view). And this box (the tooltip showing the
corresponding sentence) doesn't include the complete paragraph, so it's
easy to get lost...}''.
In contrast, DocuBurst showed text as paragraphs in a view that used
more vertical space, such that users were able to read the sentences
more easily.
``\textit{One thing I like this system is when I click some words, they
divide it as paragraph rather than the entire document...help me read
more specifically}'', said participant $Pd3$. 

When answering a given question, 7 of the 9 participants using \name\
searched or explored related information in the interface and summarized
their findings.
The remaining 2 mainly attempted to recall the answer from earlier
explorations and then referred to the interface to confirm.
For DocuBurst, this distribution was 5 participants chiefly exploring
the interface, and 4 chiefly recalling the answer.
Compared to \name, more participants using DocuBurst answered questions
from memory, almost equal to the number of participants who explored the
visualizations to find answers. 
Participant comments indicated that they felt they might spend too much
time in locating the required information.
For instance, when trying to find common concepts between two documents
(task T2), participant $Pd9$ who used DocuBurst commented that
``\textit{it is really hard to see all of them (words in the sunburst
diagram). And I really wanna expand one of those, but then I'm not
sure if it will cover all the things that I wanna see\ldots.
It's hard to go back to where you came from}''.
Similar comments were also made by those participants using DocuBurst to
first gather information before answering the question.
In this study, we did not screen participants based on document
familiarity as their responses were valuable to us regardless of their
prior knowledge of the document.
We had 1 participant in each group ($Pc9$ and $Pd5$) who had read the
document for T1.
$Pc9$ answered questions faster than the other participants, while $Pd5$
used close reading rather than Docuburst's Sunburst diagram to answer
questions, saying, ``\textit{I don't know how to use this tool to help
me read this paper.}'' 

\subsection{Task-Level Observations}
\label{sec:result_task}

We further separate task-wise participant behavior based on how they
achieved specific objectives within tasks.
This behavior was not restricted to any one task; rather, it
characterized how certain participants chose to access information
across tasks.

\textbf{Document Sensemaking:}
When exploring the full document (T1), participants across both
interfaces attempted to use the visualization to quickly get a sense of
what topics were addressed in the document.
Ten of the 18 participants (6 using \name, 4 using DocuBurst) were able
to quickly identify that the document was an ``InfoVis paper''.
Certain participant behaviors were similar across both interfaces.
Most of them explored the document using the main visualization first,
and only later resorted to close reading of the text.
Even after recognizing it as an academic paper, only 2 participants
relied on the paper structure (e.g., abstract/introduction) to get a
sense of the document.

However, DocuBurst users were more easily overwhelmed by the large
number of words in the Sunburst diagram, many of which (they felt) were
not closely related to the main theme of the document.
Participant $Pd3$ observed,  ``\textit{some words maybe appear really
frequently, but it's actually not very important ... it's just because
it's used very frequently by any document}.''
Another participant ($Pd2$) found it difficult to organize the words
into themes, saying ``\textit{it is a little bit hard to place the
information together, because you don't know what the correlation is
between (among) these things (i.e.\ the concepts provided)}''

Participants' perception of the document when using \name\ was largely
influenced by the extent of overlap between the document text and the
ontology.
For instance, the concept ``visualization'' being well-defined in the
ontology, was successfully identified by 8 out of 9 participants in T1.
However, the concept ``animation'' was not as well-defined in CSO, as a
result of which 5 out of 9 participants failed to determine that task T2
involved papers discussing animation.
In comparison, 8 of the 9 using DocuBurst were able to successfully
identify the animation theme.

\textbf{Concept Sensemaking:}
When making sense of a concept (\textbf{R2, R3}), most of the
participants chose to locate it in the main visualization first, and
only then looked at the other views to answer relevant questions.
To locate a specific concept in the visualization, participants'
strategies varied based on the solutions available in the interface and
their preference.

In \name, 5 of the 9 participants used the search feature, while the
others preferred to visually search the concept in the interface, i.e.
looking it up in the concept list or directly checking the Bubble
Treemap.
Since DocuBurst did not feature a search box available, all 9
participants set the concept to locate as the root word.
However, eight of the 9 participants failed with this strategy and had
to set alternatives of the original concept (e.g. a parent concept, a
synonym, or a substring of the target concept) as root words.
One unique strategy that at least 3 participants used to search in
DocuBurst was to start from higher-level concepts and dive deeper
towards their targets in the sunburst diagram.
Once again, their success depended on their choice of parent concepts: 
they often lost their way as they could not retrace their steps.
In comparison, participants found it more straightforward to locate
concepts in \name.

While participants using either interface chiefly attempted to define a
concept (\textbf{R1}) by referring to the context of its use
(\textbf{R2}), their approach to identify the context was different
across the interfaces.
In \name, the concordance view was used the most, with all 9
participants using this view to identify the context at least once.
This was followed by the close reading of the transcript (used by 7
participants), with the word cloud being used by 6 participants at least
once.
Although DocuBurst also provided a word cloud, only 2 participants used
it for context.
This was likely because DocuBurst's word cloud was not organized into
concepts as done in \name, and furthermore, the word cloud in
DocuBurst---designed to supplement the main visualization---only
featured proper nouns that would not otherwise be visualized in the
Sunburst diagram.
To find related concepts (\textbf{R3}), participants using \name\
chiefly referred to the Bubble Treemap while DocuBurst users referred to
the raw text view.

\textbf{Multi-document Comparison:}
We observed participants' behavior when comparing documents both at the
conceptual level and the full-text level (\textbf{R4}).
Participants using \name\ used several techniques including highlighting
concepts in the Bubble Treemap, highlighting concepts in the concept
list, checking the relevant sparklines, and comparing the word cloud
within a concept group.
Five of the 9 participants reported that these techniques were sufficient to answer all of the questions in tasks T2 and T3.
Participant $Pc1$ observed, ``\textit{just looking at this (the Bubble
Treemap for the third document in T3), you can see some colors are
different, means some different concepts exist here... you can
immediately see it}''.
When the visual clues were not enough to aid them to summarize the
similarities or differences between/among the documents, the other 4
participants resorted to close reading of the document.

In contrast, most of the participants using DocuBurst mentioned that the
visualizations and interactions were not sufficient to help them compare
the concepts or full text of the documents.
Participant $Pd5$ commented, ``\textit{the visual encoding
(distinguishing concepts between documents) is confusing to me}''.
Participant $Pd9$ felt ``\textit{it is really hard to see all of them
(concepts)}'' when they tried to identify unique concepts of one
document.
Both participant $Pd1$ and $Pd8$ were distracted by such general words
as ``\textit{part}'' and ``\textit{paper}'', because they were the only
few words marked as being shared by both documents.
As a solution, they chose to read the document text closely to make sure
their responses to the questions were accurate enough.

\subsection{Overall Feedback}
\label{sec:reslult_feedback}

\begin{figure*}[tb]
  \centering
  \includegraphics[width=\textwidth]{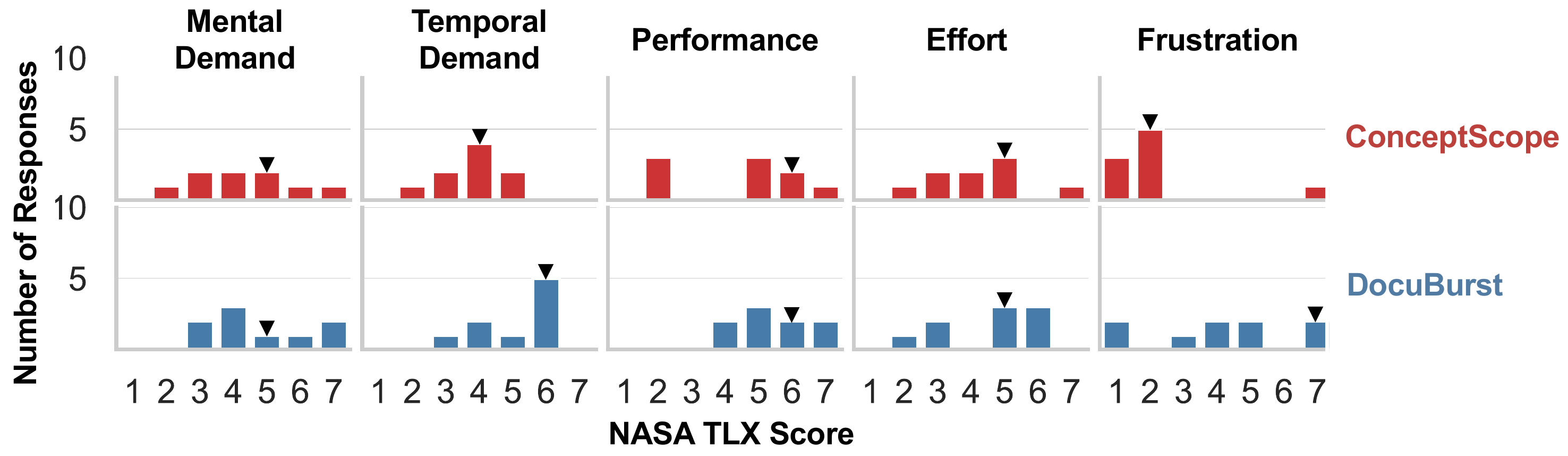}
  \caption{Distribution of NASA TLX responses showing participant
    feedback towards \name\ and DocuBurst. A $\blacktriangledown$ symbol
    indicates where a user familiar with the document ($Pc9$ for \name\
    and $Pd5$ for DocuBurst) has rated their experience.}
  \label{fig:nasa_tlx}
\end{figure*}

Fig.~\ref{fig:nasa_tlx} shows the difference in participant experience
for the study between \name\ and DocuBurst.
We can see from the figure that participants' experience was more or
less similar between the two interfaces with the exception of
frustration: participants using \name\ were less frustrated
($Md = 2, IQR = 1$) than those using DocuBurst ($Md = 4, IQR = 4$).
Observation and feedback indicated that participants using DocuBurst
found themselves distracted by less relevant concepts.
Participant $Pd3$ stated that the interface didn't provide ``important''
keywords as expected: ``\textit{When I click `person'... it (the
corresponding sector in sunburst diagram) is really big, means that it
is important. However, I don't think it is important based on what I've
seen}''.
Participant $Pd4$ mistook the document in task T1 for a medical paper
and participant $Pd9$ mistook those in T2 as related to chemistry, based
on their (mistaken) interpretation of proper nouns in the word cloud.

As general feedback, most participants using \name\ considered it
suitable to provide an overview for unfamiliar documents, while those
using DocuBurst felt it was better suited as a supplementary tool when
exploring familiar documents.
Typical comments about \name\ included ``\textit{these multiple views
are nice and easy to understand}''(participant $Pc2$), ``\textit{it
seems like a pretty useful tool especially for exploring large set of
documents to get an idea of what the main topics are, what kind of
researchers are active}'' ($Pc7$).
With DocuBurst, participant $Pd2$ suggested that ``\textit{the tool
should be used as a supplementary tool ... doesn't help too much with
understanding the document}''.
In addition, participants also reflected that the learning curves for
both tools were relatively steep.
``\textit{It was hard at the beginning, but not so hard later}'',
commented participant $Pc4$.

When regarding the features of each interface, the Level Slicer
(Sec.~\ref{sec:int_sz}) in \name\ was marked as least useful by
participants.
Participant $Pc1$ observed that ``\textit{the level slicer is probably
useful if the document is extremely complex ... but this dataset is
relatively simple}''.
Three of the participants thought the list view
(Sec.~\ref{sec:scent_leg}) was the most useful feature.
We also observed 7 participants used it for comparison tasks and 4
participants used it to search target concepts in the study.
Only 2 participants rated the Bubble Treemap (Sec.~\ref{sec:hierarchy})
as the most useful feature, while one marked it as the least useful one.
Yet, we did see 7 participants used it as their major source of visual
clues when comparing multiple documents. 
It is likely that the participants used the Bubble Treemap as providing
supplementary information to the concept list view, which they found to
be most useful.

\section{Limitations and Future Work}
Based on participant behavior and feedback, we illustrate that \name's
ontology-based visualization and grouped word clouds help participants
define and contextualize concepts, and---for a given concept---explore
other concepts related to it.
On the other hand, \name's domain dependency makes it less suitable for
reviewing text that spans multiple disciplines.
In contrast, DocuBurst's domain-agnostic reference (i.e.\ WordNet)
allows it to be applied more widely, though the overviews are less
useful when highly domain-specific content is visualized.
In addition, DocuBurst's interface is more amenable to close reading of
the document.

Our study can be considered preliminary, as we were interested in
participants' exploratory behavior, insights, and comprehension.
We plan to conduct longitudinal studies to evaluate the utility of
ConceptScope as a tool for preliminary review and further exploration
before and after close-reading of documents, and examine additional
encodings such as position constancy of concepts in the bubble treemap
for document comparison.

In the future, we plan to address issues relating to the ontology
lookup.
One main disadvantage is the dependence on ontologies that may or may
not be mature.
We currently use DBPedia to ``broaden'' our lookup, but using DBPedia
detracts from the strict definitions and relationship requirements to
which domain ontologies need to adhere.
Our Bubble Treemap visualization as well as our ontology lookup can
currently support only one ontology.
This makes it difficult to view documents of an interdisciplinary
nature.
We also intend to explore the application of our approach to real-time
visualizations of online forums or technical communication in the form
of emails or instant messengers.

\section{Conclusion}
In this paper, we proposed \name, an interface that aids a
knowledge-based exploration and comparison of documents based on a
reference domain ontology.
We present the use of a Bubble Treemap visualization as the primary
overview visualization to show the distribution of concepts for a
document of interest, and describe our approach to translate document
content into appropriate queries that best reflect the concept spread
and show their hierarchical relationships in the domain ontology.
We illustrate our approach using the Computer Science Ontology as our
reference.
We demonstrate the use of \name\ for document exploration and
comparison, and then evaluate \name\ against DocuBurst, the only other
overview visualization based on human-curated knowledge.
We find that \name\ offers greater advantages in terms of
domain-specificity, contextual views, and comparison of multiple
documents, but not for close reading of documents, or documents spanning
multiple domains.
DocuBurst's domain-agnosticism makes it more suitable for a
general-purpose document exploration tool spanning multiple domains, but
less so for multi-document comparison or in-depth, domain-specific
exploration.
Our future research aims to address this issue by enabling the use of
multiple reference ontologies, and explore text content such as online
forums and organizational communication.

\begin{acks}
We are grateful to Prof. Christopher Collins and Nathan Beals for their
help with using DocuBurst for our user study, to Sandra Bae for her
narration in our demo video, and to Oh-Hyun Kwon for his suggestions to
improve the paper.
We also thank the participants who volunteered with our study during
these challenging times and the anonymous reviewers for their feedback
and suggestions that helped improve the quality of this paper.
This research is sponsored in part by the U.S. National Science
Foundation through grant IIS-1741536.

\end{acks}

\bibliographystyle{ACM-Reference-Format}
\bibliography{references}


\begin{thebibliography}{51}


\ifx \showCODEN    \undefined \def \showCODEN     #1{\unskip}     \fi
\ifx \showDOI      \undefined \def \showDOI       #1{#1}\fi
\ifx \showISBNx    \undefined \def \showISBNx     #1{\unskip}     \fi
\ifx \showISBNxiii \undefined \def \showISBNxiii  #1{\unskip}     \fi
\ifx \showISSN     \undefined \def \showISSN      #1{\unskip}     \fi
\ifx \showLCCN     \undefined \def \showLCCN      #1{\unskip}     \fi
\ifx \shownote     \undefined \def \shownote      #1{#1}          \fi
\ifx \showarticletitle \undefined \def \showarticletitle #1{#1}   \fi
\ifx \showURL      \undefined \def \showURL       {\relax}        \fi
\providecommand\bibfield[2]{#2}
\providecommand\bibinfo[2]{#2}
\providecommand\natexlab[1]{#1}
\providecommand\showeprint[2][]{arXiv:#2}

\bibitem[\protect\citeauthoryear{Achich, Bouaziz, Algergawy, and
  Gargouri}{Achich et~al\mbox{.}}{2017}]%
        {Achich2017ontology}
\bibfield{author}{\bibinfo{person}{Nassira Achich}, \bibinfo{person}{Bassem
  Bouaziz}, \bibinfo{person}{Alsayed Algergawy}, {and} \bibinfo{person}{Faiez
  Gargouri}.} \bibinfo{year}{2017}\natexlab{}.
\newblock \showarticletitle{Ontology Visualization: An Overview}. In
  \bibinfo{booktitle}{\emph{International Conference on Intelligent Systems
  Design and Applications}}. \bibinfo{publisher}{Springer, Cham},
  \bibinfo{address}{USA}, \bibinfo{pages}{880--891}.
\newblock
\urldef\tempurl%
\url{https://doi.org/10.1007/978-3-319-76348-4_84}
\showDOI{\tempurl}


\bibitem[\protect\citeauthoryear{Alexander, Kohlmann, Valenza, Witmore, and
  Gleicher}{Alexander et~al\mbox{.}}{2014}]%
        {Alexander2014serendip}
\bibfield{author}{\bibinfo{person}{Eric Alexander}, \bibinfo{person}{Joe
  Kohlmann}, \bibinfo{person}{Robin Valenza}, \bibinfo{person}{Michael
  Witmore}, {and} \bibinfo{person}{Michael Gleicher}.}
  \bibinfo{year}{2014}\natexlab{}.
\newblock \showarticletitle{Serendip: Topic model-driven visual exploration of
  text corpora}. In \bibinfo{booktitle}{\emph{Proceedings of the IEEE
  Conference on Visual Analytics Science and Technology}}.
  \bibinfo{publisher}{IEEE}, \bibinfo{address}{USA}, \bibinfo{pages}{173--182}.
\newblock
\urldef\tempurl%
\url{https://doi.org/10.1109/vast.2014.7042493}
\showDOI{\tempurl}


\bibitem[\protect\citeauthoryear{Archambault and Purchase}{Archambault and
  Purchase}{2016}]%
        {Archambault2016can}
\bibfield{author}{\bibinfo{person}{Daniel Archambault} {and}
  \bibinfo{person}{Helen~C Purchase}.} \bibinfo{year}{2016}\natexlab{}.
\newblock \showarticletitle{Can animation support the visualisation of dynamic
  graphs?}
\newblock \bibinfo{journal}{\emph{Information Sciences}}  \bibinfo{volume}{330}
  (\bibinfo{year}{2016}), \bibinfo{pages}{495--509}.
\newblock
\urldef\tempurl%
\url{https://doi.org/10.1016/j.ins.2015.04.017}
\showDOI{\tempurl}


\bibitem[\protect\citeauthoryear{Barlow and Neville}{Barlow and
  Neville}{2001}]%
        {Barlow2001comparison}
\bibfield{author}{\bibinfo{person}{Todd Barlow} {and} \bibinfo{person}{Padraic
  Neville}.} \bibinfo{year}{2001}\natexlab{}.
\newblock \showarticletitle{A comparison of 2-D visualizations of hierarchies}.
  In \bibinfo{booktitle}{\emph{IEEE Symposium on Information Visualization}}.
  \bibinfo{publisher}{IEEE}, \bibinfo{address}{USA}, \bibinfo{pages}{131--138}.
\newblock
\urldef\tempurl%
\url{https://doi.org/10.1109/infvis.2001.963290}
\showDOI{\tempurl}


\bibitem[\protect\citeauthoryear{Bederson and Boltman}{Bederson and
  Boltman}{1999}]%
        {Bederson1999does}
\bibfield{author}{\bibinfo{person}{Benjamin~B Bederson} {and}
  \bibinfo{person}{Angela Boltman}.} \bibinfo{year}{1999}\natexlab{}.
\newblock \showarticletitle{Does animation help users build mental maps of
  spatial information?}. In \bibinfo{booktitle}{\emph{Proceedings IEEE
  Symposium on Information Visualization}}. \bibinfo{publisher}{IEEE},
  \bibinfo{address}{USA}, \bibinfo{pages}{28--35}.
\newblock
\urldef\tempurl%
\url{https://doi.org/10.1109/infvis.1999.801854}
\showDOI{\tempurl}


\bibitem[\protect\citeauthoryear{Bird, Klein, and Loper}{Bird
  et~al\mbox{.}}{2009}]%
        {Bird2009natural}
\bibfield{author}{\bibinfo{person}{Steven Bird}, \bibinfo{person}{Ewan Klein},
  {and} \bibinfo{person}{Edward Loper}.} \bibinfo{year}{2009}\natexlab{}.
\newblock \bibinfo{booktitle}{\emph{Natural language processing with Python:
  analyzing text with the natural language toolkit}}.
\newblock \bibinfo{publisher}{{O'Reilly Media, Inc.}}, \bibinfo{address}{USA}.
\newblock
\urldef\tempurl%
\url{https://doi.org/10.1007/s10579-010-9124-x}
\showDOI{\tempurl}


\bibitem[\protect\citeauthoryear{Bostrom}{Bostrom}{2015}]%
        {Bostrom2015what}
\bibfield{author}{\bibinfo{person}{Nick Bostrom}.}
  \bibinfo{year}{2015}\natexlab{}.
\newblock \bibinfo{title}{What happens when our computer gets smarter than we
  are?}
\newblock
\newblock
\urldef\tempurl%
\url{https://www.ted.com/talks/nick_bostrom_what_happens_when_our_computers_get_smarter_than_we_are?language=en}
\showURL{%
\tempurl}
\newblock
\shownote{Accessed December 3, 2019.}


\bibitem[\protect\citeauthoryear{Collins, Carpendale, and Penn}{Collins
  et~al\mbox{.}}{2009a}]%
        {Collins2009docuburst}
\bibfield{author}{\bibinfo{person}{Christopher Collins},
  \bibinfo{person}{Sheelagh Carpendale}, {and} \bibinfo{person}{Gerald Penn}.}
  \bibinfo{year}{2009}\natexlab{a}.
\newblock \showarticletitle{{DocuBurst}: Visualizing Document Content using
  Language Structure}.
\newblock \bibinfo{journal}{\emph{Computer Graphics Forum}}
  \bibinfo{volume}{28}, \bibinfo{number}{3} (\bibinfo{year}{2009}),
  \bibinfo{pages}{1039--1046}.
\newblock
\urldef\tempurl%
\url{https://doi.org/10.1111/j.1467-8659.2009.01439.x}
\showDOI{\tempurl}


\bibitem[\protect\citeauthoryear{Collins, Viegas, and Wattenberg}{Collins
  et~al\mbox{.}}{2009b}]%
        {Collins2009parallel}
\bibfield{author}{\bibinfo{person}{Christopher Collins},
  \bibinfo{person}{Fernanda~B Viegas}, {and} \bibinfo{person}{Martin
  Wattenberg}.} \bibinfo{year}{2009}\natexlab{b}.
\newblock \showarticletitle{Parallel tag clouds to explore and analyze faceted
  text corpora}. In \bibinfo{booktitle}{\emph{{IEEE} Symposium on Visual
  Analytics Science and Technology}}. \bibinfo{publisher}{IEEE},
  \bibinfo{address}{Atlantic City, NJ, USA}, \bibinfo{pages}{91--98}.
\newblock
\urldef\tempurl%
\url{https://doi.org/10.1109/VAST.2009.5333443}
\showDOI{\tempurl}


\bibitem[\protect\citeauthoryear{Cui, Zhang, Wang, Huang, Chen, Fang, Zhang,
  Lou, and Zhang}{Cui et~al\mbox{.}}{2019}]%
        {Cui2019text}
\bibfield{author}{\bibinfo{person}{Weiwei Cui}, \bibinfo{person}{Xiaoyu Zhang},
  \bibinfo{person}{Yun Wang}, \bibinfo{person}{He Huang}, \bibinfo{person}{Bei
  Chen}, \bibinfo{person}{Lei Fang}, \bibinfo{person}{Haidong Zhang},
  \bibinfo{person}{Jian-Guan Lou}, {and} \bibinfo{person}{Dongmei Zhang}.}
  \bibinfo{year}{2019}\natexlab{}.
\newblock \showarticletitle{Text-to-Viz: Automatic Generation of Infographics
  from Proportion-Related Natural Language Statements}.
\newblock \bibinfo{journal}{\emph{IEEE Transactions on Visualization and
  Computer Graphics}} \bibinfo{volume}{26}, \bibinfo{number}{1}
  (\bibinfo{year}{2019}), \bibinfo{pages}{906--916}.
\newblock
\urldef\tempurl%
\url{https://doi.org/10.1109/TVCG.2019.2934785}
\showDOI{\tempurl}


\bibitem[\protect\citeauthoryear{Dud{\'a}{\v{s}}, Lohmann, Sv{\'a}tek, and
  Pavlov}{Dud{\'a}{\v{s}} et~al\mbox{.}}{2018}]%
        {Dudas2018ontology}
\bibfield{author}{\bibinfo{person}{Marek Dud{\'a}{\v{s}}},
  \bibinfo{person}{Steffen Lohmann}, \bibinfo{person}{Vojt{\v{e}}ch
  Sv{\'a}tek}, {and} \bibinfo{person}{Dmitry Pavlov}.}
  \bibinfo{year}{2018}\natexlab{}.
\newblock \showarticletitle{Ontology visualization methods and tools: a survey
  of the state of the art}.
\newblock \bibinfo{journal}{\emph{The Knowledge Engineering Review}}
  \bibinfo{volume}{33} (\bibinfo{year}{2018}), \bibinfo{pages}{e10}.
\newblock
\urldef\tempurl%
\url{https://doi.org/10.1017/S0269888918000073}
\showDOI{\tempurl}


\bibitem[\protect\citeauthoryear{El-Assady, Gold, Acevedo, Collins, and
  Keim}{El-Assady et~al\mbox{.}}{2016}]%
        {El2016contovi}
\bibfield{author}{\bibinfo{person}{Mennatallah El-Assady},
  \bibinfo{person}{Valentin Gold}, \bibinfo{person}{Carmela Acevedo},
  \bibinfo{person}{Christopher Collins}, {and} \bibinfo{person}{Daniel Keim}.}
  \bibinfo{year}{2016}\natexlab{}.
\newblock \showarticletitle{{ConToVi:} Multi-party conversation exploration
  using topic-space views}.
\newblock \bibinfo{journal}{\emph{Computer Graphics Forum}}
  \bibinfo{volume}{35}, \bibinfo{number}{3} (\bibinfo{year}{2016}),
  \bibinfo{pages}{431--440}.
\newblock
\urldef\tempurl%
\url{https://doi.org/10.1111/cgf.12919}
\showDOI{\tempurl}


\bibitem[\protect\citeauthoryear{El-Assady, Sevastjanova, Gipp, Keim, and
  Collins}{El-Assady et~al\mbox{.}}{2017}]%
        {El2017nerex}
\bibfield{author}{\bibinfo{person}{Mennatallah El-Assady},
  \bibinfo{person}{Rita Sevastjanova}, \bibinfo{person}{Bela Gipp},
  \bibinfo{person}{Daniel Keim}, {and} \bibinfo{person}{Christopher Collins}.}
  \bibinfo{year}{2017}\natexlab{}.
\newblock \showarticletitle{NEREx: Named-Entity Relationship Exploration in
  Multi-Party Conversations}.
\newblock \bibinfo{journal}{\emph{Computer Graphics Forum}}
  \bibinfo{volume}{36}, \bibinfo{number}{3} (\bibinfo{year}{2017}),
  \bibinfo{pages}{213--225}.
\newblock
\urldef\tempurl%
\url{https://doi.org/10.1111/cgf.13181}
\showDOI{\tempurl}


\bibitem[\protect\citeauthoryear{Fellbaum}{Fellbaum}{1998}]%
        {Miller1998wordnet}
\bibfield{editor}{\bibinfo{person}{Christiane Fellbaum}} (Ed.).
  \bibinfo{year}{1998}\natexlab{}.
\newblock \bibinfo{booktitle}{\emph{WordNet: An electronic lexical database}}.
\newblock \bibinfo{publisher}{MIT press}, \bibinfo{address}{USA}.
\newblock


\bibitem[\protect\citeauthoryear{Gad, Javed, Ghani, Elmqvist, Ewing, Hampton,
  and Ramakrishnan}{Gad et~al\mbox{.}}{2015}]%
        {Gad2015themedelta}
\bibfield{author}{\bibinfo{person}{Samah Gad}, \bibinfo{person}{Waqas Javed},
  \bibinfo{person}{Sohaib Ghani}, \bibinfo{person}{Niklas Elmqvist},
  \bibinfo{person}{Tom Ewing}, \bibinfo{person}{Keith~N Hampton}, {and}
  \bibinfo{person}{Naren Ramakrishnan}.} \bibinfo{year}{2015}\natexlab{}.
\newblock \showarticletitle{ThemeDelta: dynamic segmentations over temporal
  topic models}.
\newblock \bibinfo{journal}{\emph{IEEE Transactions on Visualization and
  Computer Graphics}} \bibinfo{volume}{21}, \bibinfo{number}{5}
  (\bibinfo{year}{2015}), \bibinfo{pages}{672--685}.
\newblock
\urldef\tempurl%
\url{https://doi.org/10.1109/TVCG.2014.2388208}
\showDOI{\tempurl}


\bibitem[\protect\citeauthoryear{Glueck, Hamilton, Chevalier, Breslav, Khan,
  Wigdor, and Brudno}{Glueck et~al\mbox{.}}{2015}]%
        {Glueck2015phenoblocks}
\bibfield{author}{\bibinfo{person}{Michael Glueck}, \bibinfo{person}{Peter
  Hamilton}, \bibinfo{person}{Fanny Chevalier}, \bibinfo{person}{Simon
  Breslav}, \bibinfo{person}{Azam Khan}, \bibinfo{person}{Daniel Wigdor}, {and}
  \bibinfo{person}{Michael Brudno}.} \bibinfo{year}{2015}\natexlab{}.
\newblock \showarticletitle{PhenoBlocks: phenotype comparison visualizations}.
\newblock \bibinfo{journal}{\emph{IEEE Transactions on Visualization and
  Computer Graphics}} \bibinfo{volume}{22}, \bibinfo{number}{1}
  (\bibinfo{year}{2015}), \bibinfo{pages}{101--110}.
\newblock
\urldef\tempurl%
\url{https://doi.org/10.1109/TVCG.2015.2467733}
\showDOI{\tempurl}


\bibitem[\protect\citeauthoryear{G{\"o}rtler, Schulz, Weiskopf, and
  Deussen}{G{\"o}rtler et~al\mbox{.}}{2017}]%
        {Gortler2017bubble}
\bibfield{author}{\bibinfo{person}{Jochen G{\"o}rtler},
  \bibinfo{person}{Christoph Schulz}, \bibinfo{person}{Daniel Weiskopf}, {and}
  \bibinfo{person}{Oliver Deussen}.} \bibinfo{year}{2017}\natexlab{}.
\newblock \showarticletitle{Bubble treemaps for uncertainty visualization}.
\newblock \bibinfo{journal}{\emph{IEEE transactions on visualization and
  computer graphics}} \bibinfo{volume}{24}, \bibinfo{number}{1}
  (\bibinfo{year}{2017}), \bibinfo{pages}{719--728}.
\newblock
\urldef\tempurl%
\url{https://doi.org/10.1109/TVCG.2017.2743959}
\showDOI{\tempurl}


\bibitem[\protect\citeauthoryear{Gretarsson, O’donovan, Bostandjiev,
  H{\"o}llerer, Asuncion, Newman, and Smyth}{Gretarsson et~al\mbox{.}}{2012}]%
        {Gretarsson2012topicnets}
\bibfield{author}{\bibinfo{person}{Brynjar Gretarsson}, \bibinfo{person}{John
  O’donovan}, \bibinfo{person}{Svetlin Bostandjiev}, \bibinfo{person}{Tobias
  H{\"o}llerer}, \bibinfo{person}{Arthur Asuncion}, \bibinfo{person}{David
  Newman}, {and} \bibinfo{person}{Padhraic Smyth}.}
  \bibinfo{year}{2012}\natexlab{}.
\newblock \showarticletitle{{TopicNets:} Visual analysis of large text corpora
  with topic modeling}.
\newblock \bibinfo{journal}{\emph{ACM Transactions on Intelligent Systems and
  Technology}} \bibinfo{volume}{3}, \bibinfo{number}{2} (\bibinfo{year}{2012}),
  \bibinfo{pages}{23}.
\newblock
\urldef\tempurl%
\url{https://doi.org/10.1145/2089094.2089099}
\showDOI{\tempurl}


\bibitem[\protect\citeauthoryear{Gruber}{Gruber}{1993}]%
        {Gruber1993translation}
\bibfield{author}{\bibinfo{person}{Thomas~R. Gruber}.}
  \bibinfo{year}{1993}\natexlab{}.
\newblock \showarticletitle{A translation approach to portable ontology
  specifications}.
\newblock \bibinfo{journal}{\emph{Knowledge Acquisition}} \bibinfo{volume}{5},
  \bibinfo{number}{2} (\bibinfo{year}{1993}), \bibinfo{pages}{199--220}.
\newblock
\urldef\tempurl%
\url{https://doi.org/10.1006/knac.1993.1008}
\showDOI{\tempurl}


\bibitem[\protect\citeauthoryear{Hart and Staveland}{Hart and
  Staveland}{1988}]%
        {Hart1988development}
\bibfield{author}{\bibinfo{person}{Sandra~G Hart} {and}
  \bibinfo{person}{Lowell~E Staveland}.} \bibinfo{year}{1988}\natexlab{}.
\newblock \showarticletitle{Development of NASA-TLX (Task Load Index): Results
  of empirical and theoretical research}.
\newblock In \bibinfo{booktitle}{\emph{Advances in psychology}}.
  Vol.~\bibinfo{volume}{52}. \bibinfo{publisher}{Elsevier},
  \bibinfo{address}{USA}, \bibinfo{pages}{139--183}.
\newblock
\urldef\tempurl%
\url{https://doi.org/10.1016/S0166-4115(08)62386-9}
\showDOI{\tempurl}


\bibitem[\protect\citeauthoryear{Hellmann}{Hellmann}{2015}]%
        {DBpedialookup}
\bibfield{author}{\bibinfo{person}{Sebastian Hellmann}.}
  \bibinfo{year}{2015}\natexlab{}.
\newblock \bibinfo{title}{DBpedia lookup | DBpedia}.
\newblock
\newblock
\urldef\tempurl%
\url{https://wiki.dbpedia.org/lookup}
\showURL{%
\tempurl}
\newblock
\shownote{Accessed October 3, 2019.}


\bibitem[\protect\citeauthoryear{Henry, Fekete, and McGuffin}{Henry
  et~al\mbox{.}}{2007}]%
        {Henry2007nodetrix}
\bibfield{author}{\bibinfo{person}{Nathalie Henry},
  \bibinfo{person}{Jean-Daniel Fekete}, {and} \bibinfo{person}{Michael~J
  McGuffin}.} \bibinfo{year}{2007}\natexlab{}.
\newblock \showarticletitle{NodeTrix: a hybrid visualization of social
  networks}.
\newblock \bibinfo{journal}{\emph{IEEE transactions on visualization and
  computer graphics}} \bibinfo{volume}{13}, \bibinfo{number}{6}
  (\bibinfo{year}{2007}), \bibinfo{pages}{1302--1309}.
\newblock
\urldef\tempurl%
\url{https://doi.org/10.1109/TVCG.2007.70582}
\showDOI{\tempurl}


\bibitem[\protect\citeauthoryear{Howard}{Howard}{2014}]%
        {Howard2014wonderful}
\bibfield{author}{\bibinfo{person}{Jeremy Howard}.}
  \bibinfo{year}{2014}\natexlab{}.
\newblock \bibinfo{title}{The wonderful and terrifying implications of
  computers that can learn}.
\newblock
\newblock
\urldef\tempurl%
\url{https://www.ted.com/talks/jeremy_howard_the_wonderful_and_terrifying_implications_of_computers_that_can_learn?language=en}
\showURL{%
\tempurl}
\newblock
\shownote{Accessed December 3, 2019.}


\bibitem[\protect\citeauthoryear{J{\"a}nicke, Franzini, Cheema, and
  Scheuermann}{J{\"a}nicke et~al\mbox{.}}{2015}]%
        {Janicke2015close}
\bibfield{author}{\bibinfo{person}{Stefan J{\"a}nicke}, \bibinfo{person}{Greta
  Franzini}, \bibinfo{person}{Muhammad~Faisal Cheema}, {and}
  \bibinfo{person}{Gerik Scheuermann}.} \bibinfo{year}{2015}\natexlab{}.
\newblock \showarticletitle{On Close and Distant Reading in Digital Humanities:
  A Survey and Future Challenges.}. In \bibinfo{booktitle}{\emph{EuroVis
  (STARs)}}. \bibinfo{publisher}{The Eurographics Association},
  \bibinfo{pages}{83--103}.
\newblock
\urldef\tempurl%
\url{https://doi.org/10.2312/eurovisstar.20151113}
\showDOI{\tempurl}


\bibitem[\protect\citeauthoryear{Katifori, Halatsis, Lepouras, Vassilakis, and
  Giannopoulou}{Katifori et~al\mbox{.}}{2007}]%
        {Katifori2007ontology}
\bibfield{author}{\bibinfo{person}{Akrivi Katifori},
  \bibinfo{person}{Constantin Halatsis}, \bibinfo{person}{George Lepouras},
  \bibinfo{person}{Costas Vassilakis}, {and} \bibinfo{person}{Eugenia
  Giannopoulou}.} \bibinfo{year}{2007}\natexlab{}.
\newblock \showarticletitle{Ontology visualization methods—a survey}.
\newblock \bibinfo{journal}{\emph{Comput. Surveys}} \bibinfo{volume}{39},
  \bibinfo{number}{4} (\bibinfo{year}{2007}), \bibinfo{pages}{10}.
\newblock
\urldef\tempurl%
\url{https://doi.org/10.1145/1287620.1287621}
\showDOI{\tempurl}


\bibitem[\protect\citeauthoryear{Keim and Oelke}{Keim and Oelke}{2007}]%
        {Keim2007literature}
\bibfield{author}{\bibinfo{person}{Daniel~A Keim} {and}
  \bibinfo{person}{Daniela Oelke}.} \bibinfo{year}{2007}\natexlab{}.
\newblock \showarticletitle{Literature fingerprinting: A new method for visual
  literary analysis}. In \bibinfo{booktitle}{\emph{IEEE Symposium on Visual
  Analytics Science and Technology}}. \bibinfo{publisher}{IEEE},
  \bibinfo{address}{USA}, \bibinfo{pages}{115--122}.
\newblock
\urldef\tempurl%
\url{https://doi.org/10.1109/VAST.2007.4389004}
\showDOI{\tempurl}


\bibitem[\protect\citeauthoryear{Kovesi}{Kovesi}{2015}]%
        {Kovesi2015good}
\bibfield{author}{\bibinfo{person}{Peter Kovesi}.}
  \bibinfo{year}{2015}\natexlab{}.
\newblock \showarticletitle{Good colour maps: How to design them}.
\newblock \bibinfo{journal}{\emph{arXiv preprint arXiv:1509.03700}}
  (\bibinfo{year}{2015}).
\newblock


\bibitem[\protect\citeauthoryear{Kucher and Kerren}{Kucher and Kerren}{2015}]%
        {Kucher2015text}
\bibfield{author}{\bibinfo{person}{Kostiantyn Kucher} {and}
  \bibinfo{person}{Andreas Kerren}.} \bibinfo{year}{2015}\natexlab{}.
\newblock \showarticletitle{Text visualization techniques: Taxonomy, visual
  survey, and community insights}. In \bibinfo{booktitle}{\emph{IEEE Pacific
  Visualization Symposium}}. \bibinfo{publisher}{IEEE},
  \bibinfo{address}{Hangzhou, China}, \bibinfo{pages}{117--121}.
\newblock
\urldef\tempurl%
\url{https://doi.org/10.1109/pacificvis.2015.7156366}
\showDOI{\tempurl}


\bibitem[\protect\citeauthoryear{Lehmann, Isele, Jakob, Jentzsch, Kontokostas,
  Mendes, Hellmann, Morsey, Van~Kleef, Auer, and Bizer}{Lehmann
  et~al\mbox{.}}{2015}]%
        {Lehmann2015dbpedia}
\bibfield{author}{\bibinfo{person}{Jens Lehmann}, \bibinfo{person}{Robert
  Isele}, \bibinfo{person}{Max Jakob}, \bibinfo{person}{Anja Jentzsch},
  \bibinfo{person}{Dimitris Kontokostas}, \bibinfo{person}{Pablo~N Mendes},
  \bibinfo{person}{Sebastian Hellmann}, \bibinfo{person}{Mohamed Morsey},
  \bibinfo{person}{Patrick Van~Kleef}, \bibinfo{person}{S{\"o}ren Auer}, {and}
  \bibinfo{person}{Christian Bizer}.} \bibinfo{year}{2015}\natexlab{}.
\newblock \showarticletitle{DBpedia--a large-scale, multilingual knowledge base
  extracted from Wikipedia}.
\newblock \bibinfo{journal}{\emph{Semantic Web}} \bibinfo{volume}{6},
  \bibinfo{number}{2} (\bibinfo{year}{2015}), \bibinfo{pages}{167--195}.
\newblock
\urldef\tempurl%
\url{https://doi.org/10.3233/SW-140134}
\showDOI{\tempurl}


\bibitem[\protect\citeauthoryear{``moot'' Poole}{``moot'' Poole}{2010}]%
        {Poole2010case}
\bibfield{author}{\bibinfo{person}{Christopher ``moot'' Poole}.}
  \bibinfo{year}{2010}\natexlab{}.
\newblock \bibinfo{title}{The case for anonymity online}.
\newblock
\newblock
\urldef\tempurl%
\url{https://www.ted.com/talks/christopher_moot_poole_the_case_for_anonymity_online}
\showURL{%
\tempurl}
\newblock
\shownote{Accessed December 3, 2019.}


\bibitem[\protect\citeauthoryear{Moretti}{Moretti}{2005}]%
        {Moretti2005graphs}
\bibfield{author}{\bibinfo{person}{Franco Moretti}.}
  \bibinfo{year}{2005}\natexlab{}.
\newblock \bibinfo{booktitle}{\emph{Graphs, maps, trees: abstract models for a
  literary history}}.
\newblock \bibinfo{publisher}{Verso}, \bibinfo{address}{USA}.
\newblock


\bibitem[\protect\citeauthoryear{Munzner}{Munzner}{2009}]%
        {munzner2009nested}
\bibfield{author}{\bibinfo{person}{Tamara Munzner}.}
  \bibinfo{year}{2009}\natexlab{}.
\newblock \showarticletitle{A nested model for visualization design and
  validation}.
\newblock \bibinfo{journal}{\emph{IEEE transactions on visualization and
  computer graphics}} \bibinfo{volume}{15}, \bibinfo{number}{6}
  (\bibinfo{year}{2009}), \bibinfo{pages}{921--928}.
\newblock
\urldef\tempurl%
\url{https://doi.org/10.1109/TVCG.2009.111}
\showDOI{\tempurl}


\bibitem[\protect\citeauthoryear{Nualart and P{\'e}rez-Montoro}{Nualart and
  P{\'e}rez-Montoro}{2013}]%
        {Nualart2013texty}
\bibfield{author}{\bibinfo{person}{Jaume Nualart} {and} \bibinfo{person}{Mario
  P{\'e}rez-Montoro}.} \bibinfo{year}{2013}\natexlab{}.
\newblock \showarticletitle{Texty, a visualization tool to aid selection of
  texts from search outputs.}
\newblock \bibinfo{journal}{\emph{Information Research}} \bibinfo{volume}{18},
  \bibinfo{number}{2} (\bibinfo{year}{2013}).
\newblock


\bibitem[\protect\citeauthoryear{Oelke, Spretke, Stoffel, and Keim}{Oelke
  et~al\mbox{.}}{2011}]%
        {Oelke2011visual}
\bibfield{author}{\bibinfo{person}{Daniela Oelke}, \bibinfo{person}{David
  Spretke}, \bibinfo{person}{Andreas Stoffel}, {and} \bibinfo{person}{Daniel~A
  Keim}.} \bibinfo{year}{2011}\natexlab{}.
\newblock \showarticletitle{Visual readability analysis: How to make your
  writings easier to read}.
\newblock \bibinfo{journal}{\emph{IEEE Transactions on Visualization and
  Computer Graphics}} \bibinfo{volume}{18}, \bibinfo{number}{5}
  (\bibinfo{year}{2011}), \bibinfo{pages}{662--674}.
\newblock
\urldef\tempurl%
\url{https://doi.org/10.1109/TVCG.2011.266}
\showDOI{\tempurl}


\bibitem[\protect\citeauthoryear{Oelke, Strobelt, Rohrdantz, Gurevych, and
  Deussen}{Oelke et~al\mbox{.}}{2014}]%
        {Oelke2014comparative}
\bibfield{author}{\bibinfo{person}{Daniela Oelke}, \bibinfo{person}{Hendrik
  Strobelt}, \bibinfo{person}{Christian Rohrdantz}, \bibinfo{person}{Iryna
  Gurevych}, {and} \bibinfo{person}{Oliver Deussen}.}
  \bibinfo{year}{2014}\natexlab{}.
\newblock \showarticletitle{Comparative exploration of document collections: a
  visual analytics approach}, In \bibinfo{booktitle}{Computer Graphics Forum}.
\newblock \bibinfo{journal}{\emph{Computer Graphics Forum}}
  \bibinfo{volume}{33}, \bibinfo{number}{3}, \bibinfo{pages}{201--210}.
\newblock
\urldef\tempurl%
\url{https://doi.org/10.1111/cgf.12376}
\showDOI{\tempurl}


\bibitem[\protect\citeauthoryear{Roh, Kumara, Simpson, and Witherell}{Roh
  et~al\mbox{.}}{2016}]%
        {Roh2016ontology}
\bibfield{author}{\bibinfo{person}{Byeong-Min Roh}, \bibinfo{person}{Soundar~RT
  Kumara}, \bibinfo{person}{Timothy~W Simpson}, {and} \bibinfo{person}{P
  Witherell}.} \bibinfo{year}{2016}\natexlab{}.
\newblock \showarticletitle{Ontology-Based Laser and Thermal Metamodels for
  Metal-Based Additive Manufacturing}. In \bibinfo{booktitle}{\emph{ASME
  International Design Engineering Technical Conferences and Computers and
  Information in Engineering Conference}}. \bibinfo{publisher}{American Society
  of Mechanical Engineers (ASME)}, \bibinfo{address}{United States},
  \bibinfo{pages}{21--24}.
\newblock
\urldef\tempurl%
\url{https://doi.org/10.1115/DETC2016-60233}
\showDOI{\tempurl}


\bibitem[\protect\citeauthoryear{Salatino, Thanapalasingam, Mannocci, Osborne,
  and Motta}{Salatino et~al\mbox{.}}{2018}]%
        {Salatino2018computer}
\bibfield{author}{\bibinfo{person}{Angelo~A Salatino},
  \bibinfo{person}{Thiviyan Thanapalasingam}, \bibinfo{person}{Andrea
  Mannocci}, \bibinfo{person}{Francesco Osborne}, {and} \bibinfo{person}{Enrico
  Motta}.} \bibinfo{year}{2018}\natexlab{}.
\newblock \showarticletitle{The computer science ontology: a large-scale
  taxonomy of research areas}. In \bibinfo{booktitle}{\emph{International
  Semantic Web Conference}}. \bibinfo{publisher}{Springer International
  Publishing}, \bibinfo{address}{Cham}, \bibinfo{pages}{187--205}.
\newblock
\urldef\tempurl%
\url{https://doi.org/10.1007/978-3-030-00668-6_12}
\showDOI{\tempurl}


\bibitem[\protect\citeauthoryear{Shneiderman}{Shneiderman}{1992}]%
        {Shneiderman1992tree}
\bibfield{author}{\bibinfo{person}{Ben Shneiderman}.}
  \bibinfo{year}{1992}\natexlab{}.
\newblock \showarticletitle{Tree Visualization with Tree-Maps: 2-d
  Space-Filling Approach}.
\newblock \bibinfo{journal}{\emph{ACM Transactions on Graphics}}
  \bibinfo{volume}{11}, \bibinfo{number}{1} (\bibinfo{year}{1992}),
  \bibinfo{pages}{92--99}.
\newblock
\urldef\tempurl%
\url{https://doi.org/10.1145/102377.115768}
\showDOI{\tempurl}


\bibitem[\protect\citeauthoryear{Shneiderman}{Shneiderman}{2003}]%
        {Shneiderman2003eyes}
\bibfield{author}{\bibinfo{person}{Ben Shneiderman}.}
  \bibinfo{year}{2003}\natexlab{}.
\newblock \showarticletitle{The eyes have it: A task by data type taxonomy for
  information visualizations}.
\newblock In \bibinfo{booktitle}{\emph{The craft of information
  visualization}}. \bibinfo{publisher}{Morgan Kaufmann}, \bibinfo{address}{San
  Francisco}, \bibinfo{pages}{364--371}.
\newblock
\urldef\tempurl%
\url{https://doi.org/10.1016/B978-155860915-0/50046-9}
\showDOI{\tempurl}


\bibitem[\protect\citeauthoryear{Stasko, Catrambone, Guzdial, and
  McDonald}{Stasko et~al\mbox{.}}{2000}]%
        {Stasko2000evaluation}
\bibfield{author}{\bibinfo{person}{John Stasko}, \bibinfo{person}{Richard
  Catrambone}, \bibinfo{person}{Mark Guzdial}, {and} \bibinfo{person}{Kevin
  McDonald}.} \bibinfo{year}{2000}\natexlab{}.
\newblock \showarticletitle{An evaluation of space-filling information
  visualizations for depicting hierarchical structures}.
\newblock \bibinfo{journal}{\emph{International journal of human-computer
  studies}} \bibinfo{volume}{53}, \bibinfo{number}{5} (\bibinfo{year}{2000}),
  \bibinfo{pages}{663--694}.
\newblock
\urldef\tempurl%
\url{https://doi.org/10.1006/ijhc.2000.0420}
\showDOI{\tempurl}


\bibitem[\protect\citeauthoryear{Stasko, G{\"o}rg, and Liu}{Stasko
  et~al\mbox{.}}{2008}]%
        {Stasko2008jigsaw}
\bibfield{author}{\bibinfo{person}{John Stasko}, \bibinfo{person}{Carsten
  G{\"o}rg}, {and} \bibinfo{person}{Zhicheng Liu}.}
  \bibinfo{year}{2008}\natexlab{}.
\newblock \showarticletitle{Jigsaw: supporting investigative analysis through
  interactive visualization}.
\newblock \bibinfo{journal}{\emph{Information visualization}}
  \bibinfo{volume}{7}, \bibinfo{number}{2} (\bibinfo{year}{2008}),
  \bibinfo{pages}{118--132}.
\newblock
\urldef\tempurl%
\url{https://doi.org/10.1057/palgrave.ivs.9500180}
\showDOI{\tempurl}


\bibitem[\protect\citeauthoryear{Storey, Musen, Silva, Best, Ernst, Fergerson,
  and Noy}{Storey et~al\mbox{.}}{2001}]%
        {Storey2001jambalaya}
\bibfield{author}{\bibinfo{person}{Margaret-Anne Storey}, \bibinfo{person}{Mark
  Musen}, \bibinfo{person}{John Silva}, \bibinfo{person}{Casey Best},
  \bibinfo{person}{Neil Ernst}, \bibinfo{person}{Ray Fergerson}, {and}
  \bibinfo{person}{Natasha Noy}.} \bibinfo{year}{2001}\natexlab{}.
\newblock \showarticletitle{Jambalaya: Interactive visualization to enhance
  ontology authoring and knowledge acquisition in {Prot{\'e}g{\'e}}}. In
  \bibinfo{booktitle}{\emph{Workshop on interactive tools for knowledge
  capture}}, Vol.~\bibinfo{volume}{73}.
\newblock


\bibitem[\protect\citeauthoryear{Tufekci}{Tufekci}{2016}]%
        {Tufekci2016machine}
\bibfield{author}{\bibinfo{person}{Zeynep Tufekci}.}
  \bibinfo{year}{2016}\natexlab{}.
\newblock \bibinfo{title}{Machine intelligence makes human morals more
  important}.
\newblock
\newblock
\urldef\tempurl%
\url{https://www.ted.com/talks/zeynep_tufekci_machine_intelligence_makes_human_morals_more_important/transcript?referrer=playlist-talks_on_artificial_intelligen#t-3550}
\showURL{%
\tempurl}
\newblock
\shownote{Accessed October, 2020.}


\bibitem[\protect\citeauthoryear{Van~Ham, Wattenberg, and Vi{\'e}gas}{Van~Ham
  et~al\mbox{.}}{2009}]%
        {Van2009mapping}
\bibfield{author}{\bibinfo{person}{Frank Van~Ham}, \bibinfo{person}{Martin
  Wattenberg}, {and} \bibinfo{person}{Fernanda~B Vi{\'e}gas}.}
  \bibinfo{year}{2009}\natexlab{}.
\newblock \showarticletitle{Mapping text with phrase nets}.
\newblock \bibinfo{journal}{\emph{IEEE Transactions on Visualization and
  Computer Graphics}} \bibinfo{volume}{15}, \bibinfo{number}{6}
  (\bibinfo{year}{2009}), \bibinfo{pages}{1169--1176}.
\newblock
\urldef\tempurl%
\url{https://doi.org/10.1109/TVCG.2009.165}
\showDOI{\tempurl}


\bibitem[\protect\citeauthoryear{Vi\'{e}gas, Wattenberg, and
  Feinberg}{Vi\'{e}gas et~al\mbox{.}}{2009}]%
        {Viegas2009participatory}
\bibfield{author}{\bibinfo{person}{Fernanda~B Vi\'{e}gas},
  \bibinfo{person}{Martin Wattenberg}, {and} \bibinfo{person}{Jonathan
  Feinberg}.} \bibinfo{year}{2009}\natexlab{}.
\newblock \showarticletitle{Participatory visualization with wordle}.
\newblock \bibinfo{journal}{\emph{IEEE transactions on visualization and
  computer graphics}} \bibinfo{volume}{15}, \bibinfo{number}{6}
  (\bibinfo{year}{2009}), \bibinfo{pages}{1137--1144}.
\newblock
\urldef\tempurl%
\url{https://doi.org/10.1109/TVCG.2009.171}
\showDOI{\tempurl}


\bibitem[\protect\citeauthoryear{w3.org}{w3.org}{2013}]%
        {Sparql}
\bibfield{author}{\bibinfo{person}{w3.org}.} \bibinfo{year}{2013}\natexlab{}.
\newblock \bibinfo{title}{SPARQL 1.1 Query Language}.
\newblock
\newblock
\urldef\tempurl%
\url{https://www.w3.org/TR/sparql11-query/}
\showURL{%
\tempurl}
\newblock
\shownote{Accessed June 9, 2019.}


\bibitem[\protect\citeauthoryear{Wang, Wang, Dai, and Wang}{Wang
  et~al\mbox{.}}{2006}]%
        {Wang2006visualization}
\bibfield{author}{\bibinfo{person}{Weixin Wang}, \bibinfo{person}{Hui Wang},
  \bibinfo{person}{Guozhong Dai}, {and} \bibinfo{person}{Hongan Wang}.}
  \bibinfo{year}{2006}\natexlab{}.
\newblock \showarticletitle{Visualization of large hierarchical data by circle
  packing}. In \bibinfo{booktitle}{\emph{Proceedings of the ACM CHI conference
  on Human Factors in computing systems}}. \bibinfo{publisher}{Association for
  Computing Machinery}, \bibinfo{address}{New York, NY, United States},
  \bibinfo{pages}{517--520}.
\newblock
\urldef\tempurl%
\url{https://doi.org/10.1145/1124772.1124851}
\showDOI{\tempurl}


\bibitem[\protect\citeauthoryear{Wattenberg and Vi{\'e}gas}{Wattenberg and
  Vi{\'e}gas}{2008}]%
        {Wattenberg2008wordtree}
\bibfield{author}{\bibinfo{person}{Martin Wattenberg} {and}
  \bibinfo{person}{Fernanda~B. Vi{\'e}gas}.} \bibinfo{year}{2008}\natexlab{}.
\newblock \showarticletitle{The {Word Tree}, an Interactive Visual
  Concordance}.
\newblock \bibinfo{journal}{\emph{IEEE Transactions on Visualization and
  Computer Graphics}} \bibinfo{volume}{14}, \bibinfo{number}{6}
  (\bibinfo{year}{2008}), \bibinfo{pages}{1221--1228}.
\newblock
\urldef\tempurl%
\url{https://doi.org/10.1109/TVCG.2008.172}
\showDOI{\tempurl}


\bibitem[\protect\citeauthoryear{Wei, Liu, Song, Pan, Zhou, Qian, Shi, Tan, and
  Zhang}{Wei et~al\mbox{.}}{2010}]%
        {Wei2010tiara}
\bibfield{author}{\bibinfo{person}{Furu Wei}, \bibinfo{person}{Shixia Liu},
  \bibinfo{person}{Yangqiu Song}, \bibinfo{person}{Shimei Pan},
  \bibinfo{person}{Michelle~X Zhou}, \bibinfo{person}{Weihong Qian},
  \bibinfo{person}{Lei Shi}, \bibinfo{person}{Li Tan}, {and}
  \bibinfo{person}{Qiang Zhang}.} \bibinfo{year}{2010}\natexlab{}.
\newblock \showarticletitle{Tiara: a visual exploratory text analytic system}.
  In \bibinfo{booktitle}{\emph{Proceedings of the ACM International Conference
  on Knowledge Discovery and Data Mining}}. \bibinfo{publisher}{Association for
  Computing Machinery}, \bibinfo{address}{New York, NY, United States},
  \bibinfo{pages}{153--162}.
\newblock
\urldef\tempurl%
\url{https://doi.org/10.1145/1835804.1835827}
\showDOI{\tempurl}


\bibitem[\protect\citeauthoryear{Willett, Heer, and Agrawala}{Willett
  et~al\mbox{.}}{2007}]%
        {Willett2007scented}
\bibfield{author}{\bibinfo{person}{Wesley Willett}, \bibinfo{person}{Jeffrey
  Heer}, {and} \bibinfo{person}{Maneesh Agrawala}.}
  \bibinfo{year}{2007}\natexlab{}.
\newblock \showarticletitle{Scented widgets: Improving navigation cues with
  embedded visualizations}.
\newblock \bibinfo{journal}{\emph{IEEE Transactions on Visualization and
  Computer Graphics}} \bibinfo{volume}{13}, \bibinfo{number}{6}
  (\bibinfo{year}{2007}), \bibinfo{pages}{1129--1136}.
\newblock
\urldef\tempurl%
\url{https://doi.org/10.1109/TVCG.2007.70589}
\showDOI{\tempurl}


\bibitem[\protect\citeauthoryear{Witherell, Krishnamurty, and Grosse}{Witherell
  et~al\mbox{.}}{2007}]%
        {Witherell2007ontologies}
\bibfield{author}{\bibinfo{person}{Paul Witherell}, \bibinfo{person}{Sundar
  Krishnamurty}, {and} \bibinfo{person}{Ian~R Grosse}.}
  \bibinfo{year}{2007}\natexlab{}.
\newblock \showarticletitle{Ontologies for supporting engineering design
  optimization}.
\newblock \bibinfo{journal}{\emph{Journal of Computing and Information Science
  in Engineering}} \bibinfo{volume}{7}, \bibinfo{number}{2}
  (\bibinfo{year}{2007}), \bibinfo{pages}{141--150}.
\newblock
\urldef\tempurl%
\url{https://doi.org/10.1115/1.2720882}
\showDOI{\tempurl}


\end{thebibliography}

\end{document}